\documentclass[twocolumn]{aastex62}
\usepackage{times}
\usepackage{amsmath}
\usepackage{textcomp}
\usepackage{amssymb}
\usepackage[flushleft]{threeparttable}
\usepackage{boldline}
\usepackage{tabularx}
\usepackage{lineno}
\usepackage{gensymb}
\usepackage{url}
\usepackage{afterpage}
\usepackage{float}

\newcommand{\phflux}{\mbox{${\rm \, ph \,\, cm^{-2} \, s^{-1}}$}}
\newcommand{\ergflux}{\mbox{${\rm \, erg \,\, cm^{-2} \, s^{-1}}$}}
\newcommand{\gm}{$\gamma$}

\shorttitle{logN-logS}
\shortauthors{Marcotulli et al.}

\begin{document}

\title{{\bf Source-count Distribution of Gamma-Ray Blazars}}

\author{L. Marcotulli}
\affil{Department of Physics and Astronomy, Clemson University, Kinard Lab of Physics, Clemson, SC 29634-0978, USA}
\author{M. Di Mauro}
\affil{NASA Goddard Space Flight Center, Greenbelt, MD 20771, USA}
\affil{Catholic University of America, Department of Physics, Washington DC 20064, USA}
\author{M. Ajello}
\affil{Department of Physics and Astronomy, Clemson University, Kinard Lab of Physics, Clemson, SC 29634-0978, USA}

\email{lmarcot@g.clemson.edu}

\begin{abstract}
	With ten years of operation and an exceptional dataset, the Fermi-Large Area Telescope allows us to unveil the detailed composition of the extragalactic \gm-ray sky above $\rm 100\,MeV$. In this paper, we derive the intrinsic source-count distribution ({\it logN-logS}) of extragalactic sources (i.e., blazars) at $|b|>20\degree$ via the {\it efficiency correction} method. With this approach, we are able to measure the distribution down to a photon flux of $\sim10^{-10} \phflux$ and to an energy flux of $\sim10^{-12}$\ergflux. In both cases the {\it logN-logS} becomes flatter at low fluxes. Moreover, we show that this {\it logN-logS} is representative of the blazar population (assuming the majority of unassociated sources are blazars) and allows us to constrain its evolution quite effectively. Among recently proposed evolutionary models, we find that the Pure Density Evolution (PDE) model best describes the evolutionary properties of the blazar population and that their integrated emission accounts for $\sim 50^{+10}_{-5}\%$ of the total extragalactic \gm-ray background.
\end{abstract}

\section{Introduction}\label{sec:intro}
Understanding the composition of the Extragalactic \gm-ray Background (EGB) is key to untangle the origin of this non-thermal radiation field. Since its first detection by the second Small Astronomy Satellite (SAS-2, \citealp[][]{1975ApJ...198..163F}), scientists have been investigating the mystery of the Universe's \gm-ray glow.
Most recently, using data collected by the Large Area Telescope (LAT, \citealp{2009ApJ...697.1071A}) onboard the Fermi Gamma-ray Space Telescope, \citet{2015ApJ...799...86A} have precisely measured the EGB between $\rm 100\,MeV$ and $\rm 820\,GeV$, for which they found an integrated intensity of $(1.13\pm0.07)\times 10^{-5}{\rm \, ph \,\, cm^{-2} \, s^{-1} \, sr^{-1}}$. 

The EGB is produced by three components: resolved sources, unresolved sources, and truly diffuse processes. The most numerous source class resolved by the LAT is that of blazars (i.e.,\ active galactic nuclei, AGNs, with relativistic jets pointing toward the observer, at a viewing angle, $\theta_{\rm v}\lesssim10\degree$, \citealp[see e.g.,][]{2015ApJ...810...14A}). Indeed, due to relativistic beaming, the bulk of their radiation falls in the \gm-ray energy range, making them extremely bright at these frequencies. Other LAT-detected populations include misaligned AGNs (MAGNs, i.e.,\ AGNs with jets pointing at $\theta_{\rm v}\gtrsim10\degree$, \citealp[see][]{2010ApJ...720..912A}),  star-forming galaxies (SFGs, i.e.,\ galaxies whose \gm-ray emission is powered by star-formation activity, \citealp[see][]{2012ApJ...755..164A}), and narrow-line Seyfert 1 galaxies \citep[NLSy1s, see][]{2018ApJ...853L...2P}. These sources (particularly MAGNs and SFGs) are fainter than blazars in \gm~rays, although they are much more numerous. Hence, although very few have been detected by the LAT so far, they have been found to significantly contribute to the 
unresolved part of the EGB, referred to as the isotropic diffuse \gm-ray background (IGRB, \citealp{2015ApJ...799...86A}; see also \citealp{2011ApJ...733...66I, 2014ApJ...786...40L, 2014ApJ...780..161D,2015ApJ...800L..27A}).
Additionally, the IGRB may also contain the emission of truly diffuse processes (see \citealp{2015PhR...598....1F} for a review), such as dark matter (DM) annihilation \citep[e.g.,][]{2001PhRvL..87y1301B,2007PhRvD..76b3517A,2015PhRvD..91l3001D}. 

Most studies which aim at resolving the different contribution to the EGB focus primarily
on the contribution of the point source populations.
In the case of star-forming galaxies and misaligned AGNs, due to the paucity of data, these studies rely either on empirical relations between luminosity functions obtained in different wavelengths \citep[see][]{2010ApJ...722L.199F,2011ApJ...733...66I,2013ApJ...773..104C,2014ApJ...780..161D} or on cross-correlation of LAT data with catalogs of known sources \citep[see][]{2009MNRAS.400.2122A,2017ApJS..232...10C,2018PhRvD..98j3007A}. Instead, it is possible to directly determine the intrinsic source-count distribution of blazars (i.e.,\ their distribution in flux, usually referred to as {\it logN-logS}) from available data using different methods \citep[see e.g.,][]{1992ApJ...399..345E,2010ApJ...720..435A,2012ApJ...753...45S,2016PhRvL.116o1105A,2016ApJS..225...18Z,2016ApJ...832..117L,2018ApJ...856..106D}.
Previous results have reported that blazars can only account for $\sim 50^{+12}_{-11}\%$ of the total EGB \citep[see][]{2015ApJ...800L..27A} and, importantly, are not able to explain the IGRB below 100 GeV. {However, by} taking into account the integrated emission from MAGNs and SFGs, it has been found that these three populations can naturally resolve the total EGB intensity, leaving little or no room for diffuse processes interpretations \citep[][]{2015ApJ...800L..27A, 2015PhRvD..91l3001D}. 

The upgraded Pass 8 dataset \citep{2013arXiv1303.3514A} increases the effective area, in particular with high impact below a few hundred MeV, and improves the point-spread function (PSF) and energy resolution of the LAT across all energy ranges. Moreover, the LAT has been in orbit for more than 10 years, scanning the entire \gm-ray sky every 3 hours, providing an extremely large amount of data. In combination with the exceptional quality of the Pass 8 dataset, this enables the precise characterization of the intrinsic population of LAT resolved sources with unprecedented accuracy. 
In this paper, we present an improved study to estimate the contribution of resolved point sources (i.e.,\ blazars\footnote{
Blazars constitute 95\% of the LAT-detected sources above Galactic latitudes ($|b|>20\degree$). In Section~\ref{sec:complete}, using a reasonable zeroth-order assumption that this ratio holds true for the unresolved point sources, we construct the {\it logN-logS} of blazars and show that the one derived in this work is representative of the blazar population. Throughout the paper we will therefore refer to the `blazar {\it logN-logS}', anticipating this result.}) to the EGB, from $\rm 100\,MeV$ up to $\rm 1\,TeV$. We use eight years of Pass 8 data in our analysis and we employ the efficiency correction method in order to derive the intrinsic blazars {\it logN-logS} \citep{2010ApJ...720..435A,2016PhRvL.116o1105A,2018ApJ...856..106D}. This also allows us to constrain their evolution models. 

The paper is organized as follows. In Section~\ref{sec:analysismethod} we describe the {\it efficiency correction} method, along with our data selection (Section~\ref{sec:data}) and detection pipeline (Section~\ref{sec:pipe}). We report the results for the real and simulated sky in Sections~\ref{sec:real} and \ref{sec:sim}, respectively. In Section~\ref{sec:eff} the detection efficiency and {\it logN-logS} are derived. In Section~\ref{sec:mlfit} we lay out the maximum likelihood fit applied to the {\it logN-logS}, in Section~\ref{sec:syst} we detail the systematics of the analysis, and in Section~\ref{sec:complete} we demonstrate that the derived {\it logN-logS} is representative of the blazar population. In Section~\ref{sec:disc} we derive the contribution of blazars to the EGB and compare our results to the predictions of blazars' evolution models.

\section{Analysis Method}\label{sec:analysismethod}
The aim of this work is to calculate the intrinsic source-count distribution of point sources in the extragalactic \gm-ray sky.
According to all the LAT catalogs, the majority of the extragalactic sources are blazars,
while SFGs and MAGNs are observed in small numbers \citep[$\sim$10-30, see][]{2010ApJS..188..405A, 2012ApJS..199...31N, 2015ApJS..218...23A,2020ApJS..247...33A}.
Due to the low statistics the intrinsic distribution of the latter cannot rely on the \gm-ray observations alone, but requires information gained through other energy bands, leading to large uncertainties.
On the contrary, it is possible to derive the contribution of blazars to the EGB directly using the available LAT data. A technique that has only a small dependency on extrapolation, and therefore produces low uncertainties, is the {\it efficiency correction} method \citep{2010ApJ...720..435A,2016PhRvL.116o1105A,2018ApJ...856..106D}.
By means of exhaustive Monte Carlo simulations it is possible to determine the biases of the survey and analysis and to derive the blazars' intrinsic source-count distribution.

The main steps of the {\it efficiency correction} method are the following:
\begin{enumerate}
        \item Analyze LAT data in order to detect point sources in the real sky (we will refer to the obtained catalog as the {\it real} catalog) and derive their flux distribution (see Section~\ref{sec:real})
        \item Generate different realizations of the extragalactic sky via Monte Carlo simulations, which include (among other ingredients) an isotropically distributed source population with the (spectral and flux) characteristics of blazars. 
        \item Analyze each realization and detect point sources adopting the same procedure as for the real sky (see step 1). 
        \item Derive the detection efficiency ($\omega(S)$, Section~\ref{sec:eff}), i.e.,\ the probability of detecting a source within a given flux and the surveyed solid angle as a function of its flux, by comparing (for each simulation) the sources which have been detected and the sources which were actually simulated.
        \item Use the detection efficiency to correct the real catalog derived in step 1 (above) to obtain the intrinsic source-count distribution of extragalactic sources ({\it logN-logS}, Section~\ref{sec:eff}).
\end{enumerate}
\subsection{Data Selection}\label{sec:data}
We consider 8 years of LAT data, starting from 2008 August 4 (U.T. 15:43:36.00) to 2016 August 2 (U.T. 05:44:11.99). We apply a cut on Galactic latitude, $|{b}|> 20\degree$, to exclude Galactic sources and reduce contamination from the diffuse Galactic emission\footnote{The most recent release of the Galactic diffuse model tuned on Pass 8 (gll\_iem\_v07) was not available at the time of the analysis. Hence, the cut on Galactic latitude was chosen to minimize the uncertainties related to the bright Galactic diffuse emission and its modeling.}. For the analysis, we adopt the most recent release of the Pass 8 data set (P8R3, \citealp{2018arXiv181011394B}), covering the energy range from $\rm 100\,MeV$ up to $\rm 1\,TeV$. As we are interested in point source detection, we use events belonging to the $\rm SOURCE$ event class, with the corresponding instrument response functions, $\rm P8R3\_SOURCE\_V2$.
For the event type selection, we adopt a component-wise analysis, following the selections used for the preliminary LAT 8-year Point Source List (FL8Y\footnote{\url{https://fermi.gsfc.nasa.gov/ssc/data/access/lat/fl8y}}$^{,}$\footnote{Our analysis preceded the release of the fourth Fermi-LAT source catalog (4FGL, \citealp{2020ApJS..247...33A}) and therefore has been calibrated with respect to the FL8Y. All comparisons have therefore been made with the FL8Y throughout the text. We note that the 4FGL uses the same PSF and zenith angle selection as the FL8Y, but includes improved templates for the Galactic diffuse emissions. The 4FGL is now available at \url{https://fermi.gsfc.nasa.gov/ssc/data/access/lat/8yr_catalog/}.}). Depending on the quality of the angular reconstruction of the events, the Pass 8 dataset characterizes the photons in PSF event types, from PSF0 to PSF3, where PSF0 has the worst and PSF3 the best direction reconstruction. 
To minimize the contribution of low-energy Earth limb emission, while maximizing the statistics and the direction reconstruction quality of the data sample, we use PSF2 and PSF3 for the energy range $\rm 100\,MeV - 300\,MeV$; PSF1, PSF2 and PSF3 for the energy range $\rm 300\,MeV - 1\,GeV$; and finally all PSF event types above 1 GeV. The maximum zenith angles considered for the three components are $90\degree$, $100\degree$ and $105\degree$, respectively.
\begin{figure*}[t!]
\centering
	\includegraphics[width=\columnwidth]{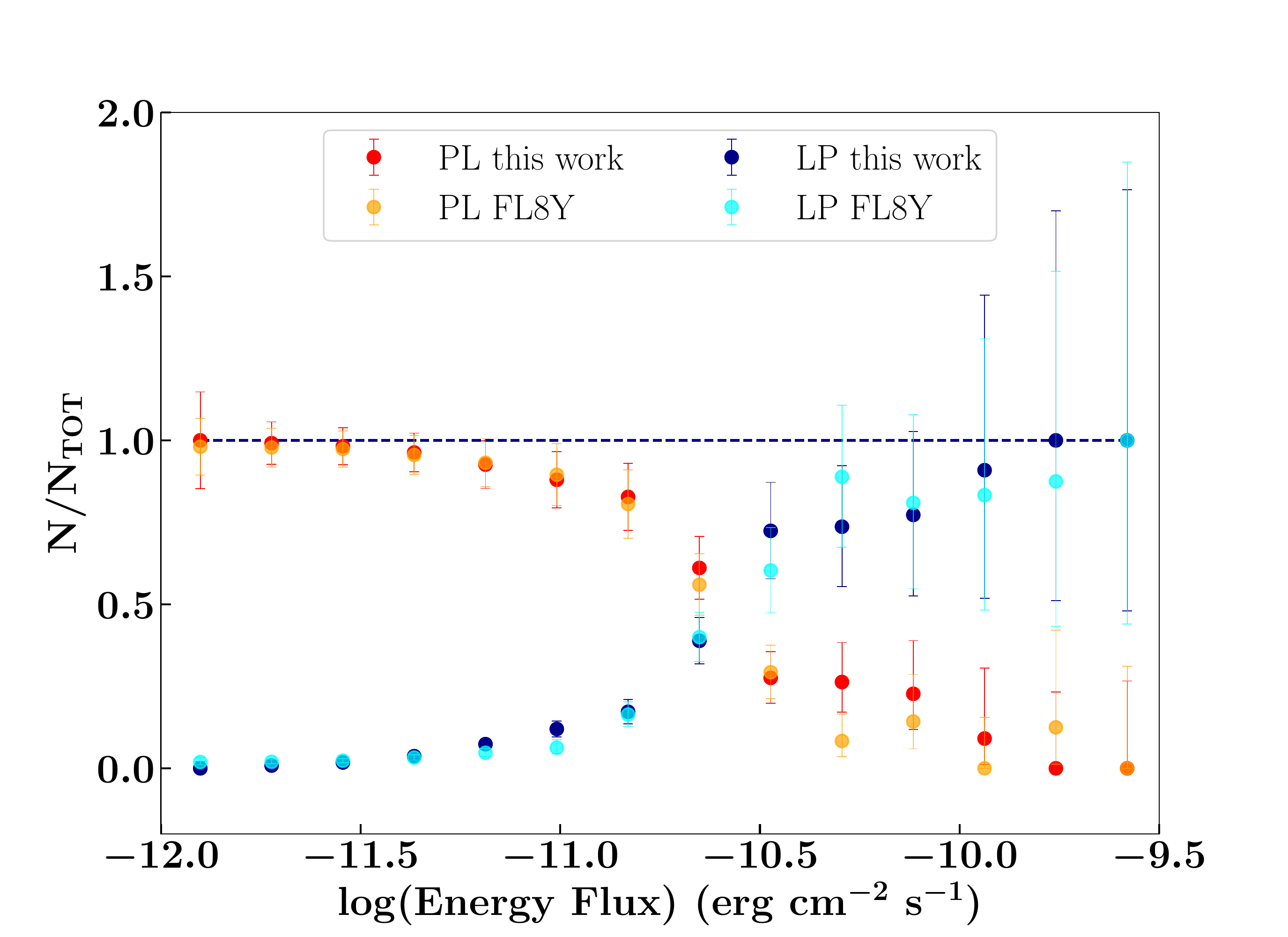}
        \includegraphics[width=0.48\textwidth]{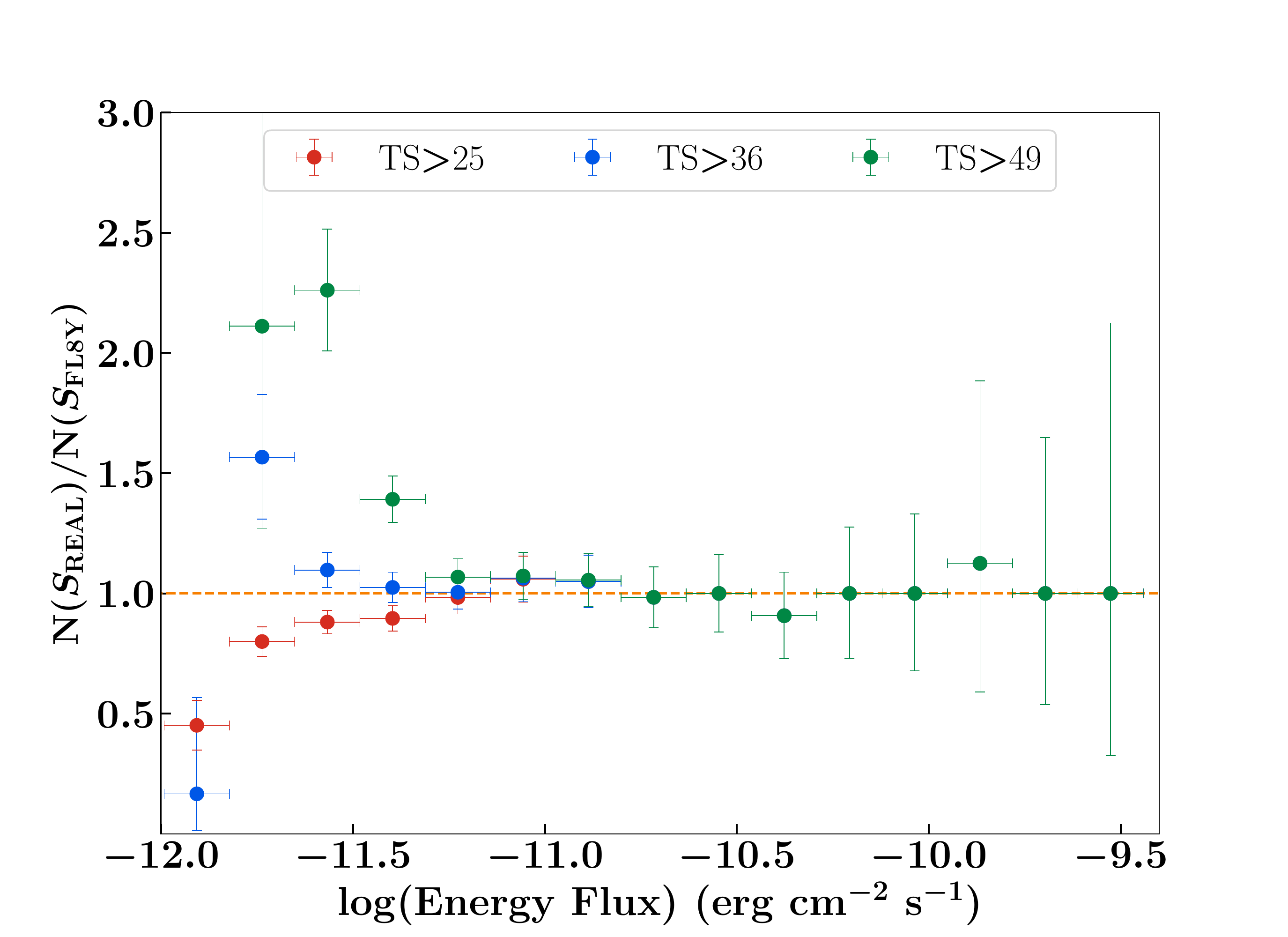}
	\caption{{\bf Left panel:} Relative number of sources with power-law spectral models and sources with log-parabolic spectral models in our catalog and the FL8Y as a function of energy flux. The distributions of both power laws and log-parabolas are in very good agreement between the two catalogs, demonstrating consistence with the FL8Y results. {\bf Right panel:} Ratio between the distribution of sources in detected energy flux for our catalog (N($S_{\rm REAL}$)) and the FL8Y (N($S_{\rm FL8Y}$)) as a function of $TS$. The ratios are compatible with one at every $TS$ for higher fluxes. The largest discrepancy appears at the lowest flux values ($\log(S)<-11.3\,\rm erg~cm^{-2}~s^{-1}$), which is expected as fainter sources are harder to detect and the two detection pipelines are different.\label{fig:relnumbers}}
\end{figure*}
In order to account for the diffuse emission, both the Galactic interstellar emission model (IEM) and isotropic diffuse background model are included in the analysis (see Section~\ref{sec:pipe} for the details on data analysis). For both, we employ standard templates released with Pass 8: gll\_iem\_v06 \citep{2016ApJS..223...26A} and iso\_P8R3\_SOURCE\_V2\footnote{\url{https://fermi.gsfc.nasa.gov/ssc/data/access/lat/BackgroundModels.html}}. To assess the systematic uncertainty due to imperfections of the IEM (see Section~\ref{sec:syst}), we also perform the analysis using an alternative template employed for the study of the Galactic center \citep[see][]{2017ApJ...840...43A}. 

\subsection{Detection pipeline}\label{sec:pipe}
Throughout the analysis we adopt the \texttt{fermipy}\footnote{http://fermipy.readthedocs.io/en/latest/} package \citep{2017ICRC...35..824W}, which is a wrapper around the Fermi-\texttt{ScienceTools} that automates the LAT data analysis.
The detection pipeline for both real and simulated data is the same. 
The extragalactic \gm-ray sky ($|\rm b|>20\degree$) is divided into regions of interest (ROIs) of 15\degree$\times$15\degree, uniformly spaced in longitude, for a total of 360 ROIs. An overlap of 3\degree~is kept between adjacent ROIs in order to accurately characterize sources at the edges. We binned our data with a pixel size of 0.1\degree, considering 8 energy bins per decade\footnote{The energy bins considered are: 100-131\,MeV, 131-173\,MeV, 173-227\,MeV, 227-300\,MeV, 300-405\,MeV, 405-547\,MeV, 547-740\,MeV, 740-1000\,MeV, 1-1.34\,GeV, 1.34-1.79\,GeV, 1.79-2.41\,GeV, 2.41-3.23\,GeV, 3.23-4.33\,GeV, 4.33-5.81\,GeV, 5.81-7.79\,GeV, 7.79-10.4\,GeV, 10.4-14.0\,GeV, 14.0-18.7\,GeV, 18.7-25.1\,GeV, 25.1-33.7\,GeV, 33.7-45.2\,GeV, 45.2-60.6\,GeV, 60.6-81.3\,GeV, 81.3-109\,GeV, 109-146\,GeV, 146-196\,GeV, 196-262\,GeV, 262-352\,GeV, 352-472\,GeV, 472-633\,GeV, 633-850\,GeV, 850-1000\,GeV.}.
The initial model for each ROI contains the isotropic template, whose normalization is free to vary, and the Galactic IEM, whose normalization and spectral index are free to vary. In each ROI, sources are detected blindly; i.e., no input catalog has been used in this work. We enable energy dispersion for the sources. 
 
The analysis is initialized using the standard Fermi-\texttt{ScienceTools} (\texttt{gtselect}, \texttt{gtmktime}, \texttt{gtmkcube}, \texttt{gtexpcube}, \texttt{gtsrcmap}).
We subsequently employ a maximum likelihood algorithm (\texttt{find\_sources}) that generates the Test Statistic\footnote{The $TS$ is defined as twice the difference of the log-likelihood between the test (presence of the source) and the null hypothesis: $TS = 2\times(\rm log \mathcal{L}_{\rm TEST}-\rm log \mathcal{L}_{\rm NULL})$, which is also known as likelihood ratio test \citep{neymanpeirson1933,wilks1938}. The $TS$ is connected to the significance of source detection, $\sigma=\sqrt {TS}$ \citep[valid for one degree of freedom,][]{1996ApJ...461..396M}.} ({\it TS}) map of the ROI, scans it to identify the significant peaks, then adds sources to the model centered at the peak positions. The algorithm is run iteratively in order to first detect and add the most significant sources ($TS>64$), requiring a minimum angular separation of 0.4\degree. 
Afterwards the analysis is repeated for $TS>36$ and minimum angular separation 0.3\degree, and lastly for $TS>20$ and minimum angular separation 0.2\degree~(minimum angular separation at which two point sources with TS=20 can be distinguished).
\begin{figure*}[t]
\centering
\includegraphics[width=0.48\textwidth]{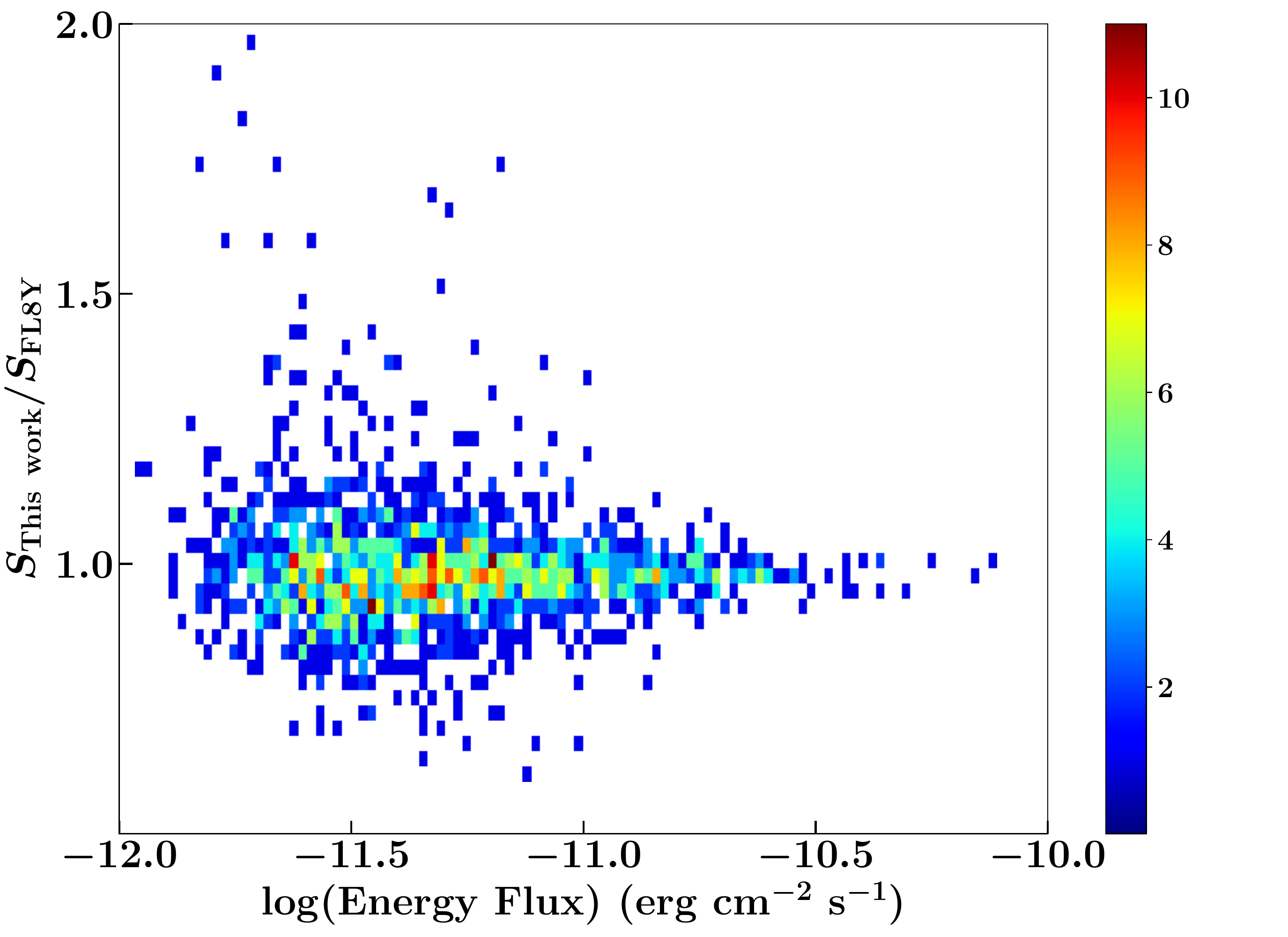}\
\includegraphics[width=0.48\textwidth]{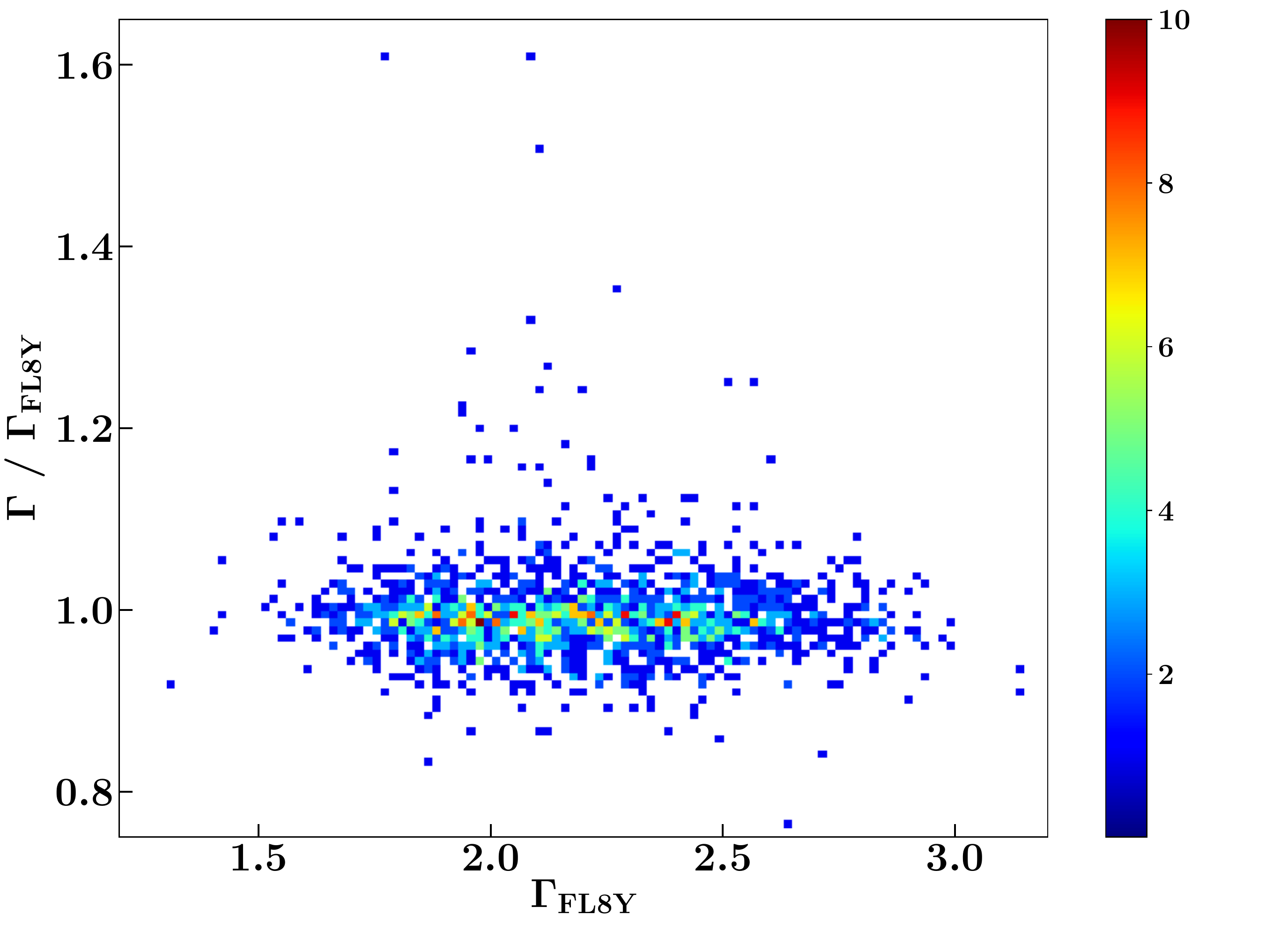}\
\includegraphics[width=0.48\textwidth]{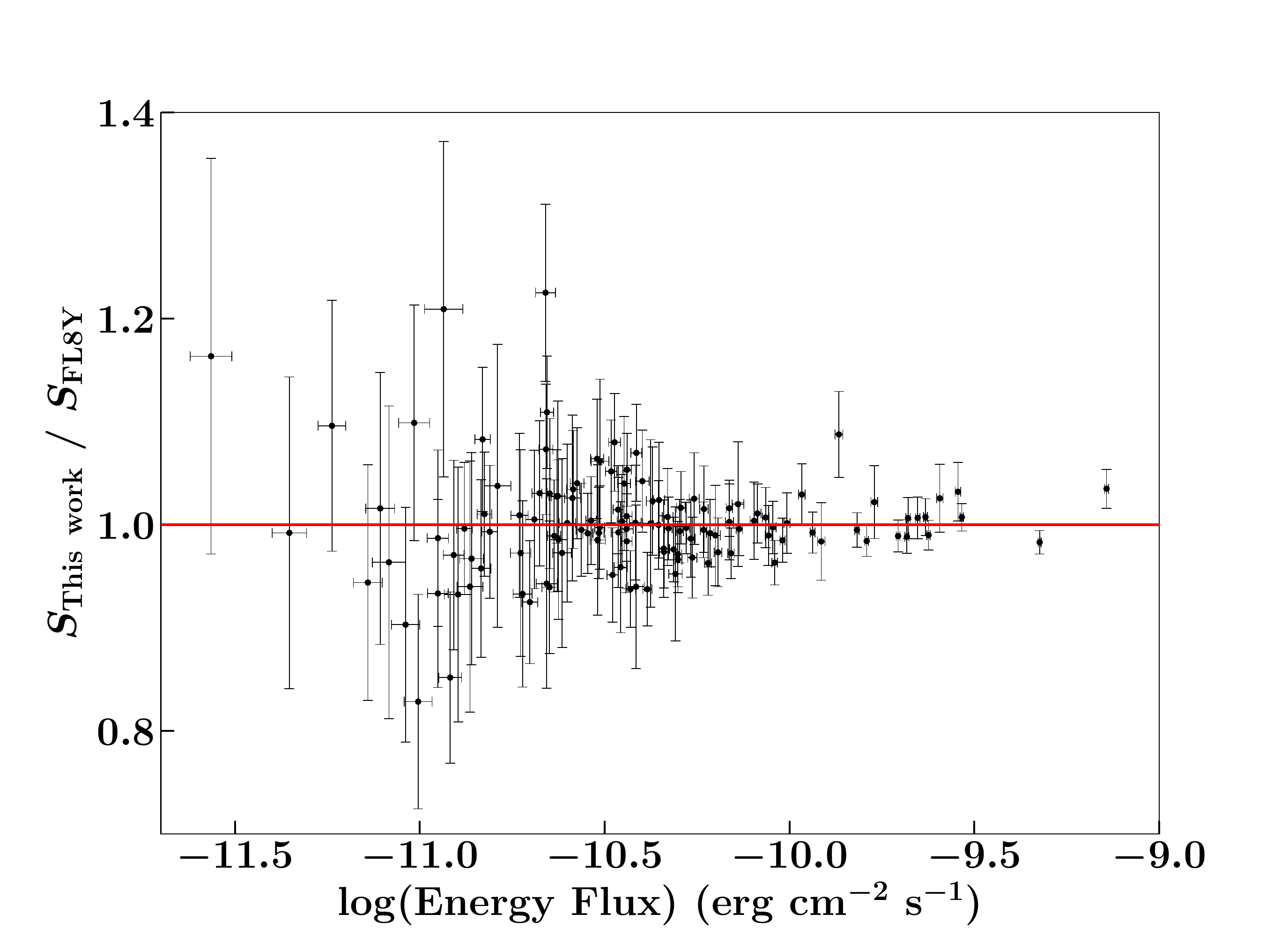}\
\includegraphics[width=0.48\textwidth]{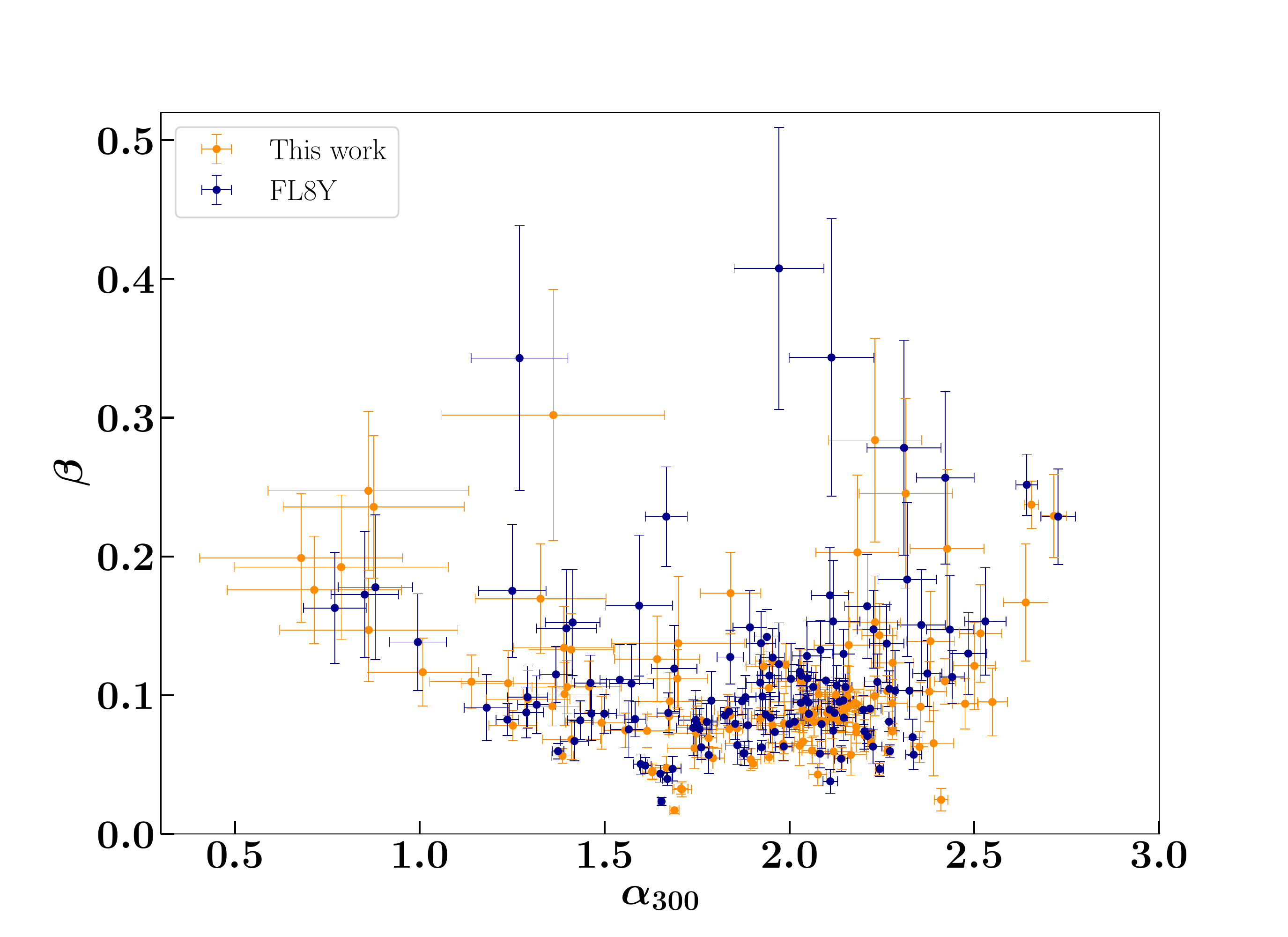}\
	\caption{{\bf Left panels:} Energy flux comparison between our detected catalog ($S_{\rm This~work}$) and the FL8Y ($S_{\rm FL8Y}$) sources with positions that match within 95\% error. The top panel is for sources whose best-fit is a power-law shape in both catalogs (2d-histogram), and the bottom one is for the log-parabolas (the red line is the one-to-one correlation). As can be seen, for both catalogs the fluxes match reasonably well, within errors. In fact, the ratios are all consistent with one. It can be seen that for power laws at lower flux values the spread with respect to the one-to-one correlation line is larger, but still symmetric and consistent with one considering the statistical errors. {\bf Right panels:} Comparison between spectral indices of matching power laws (2d-histogram, top), and $\alpha_{300}$ and $\beta$ for log-parabolas (bottom). The power-law indices are in very good agreement as the ratios are all concentrated around the correlation line. For the log-parabolas, the parameters $\alpha_{300}$ and $\beta$ are in good agreement between the catalogs. \label{fig:ourvs4fgl}}
\end{figure*}
The sources are initially considered to have power-law spectra. Then, since many blazars are known to have curved spectra, we test the most significant ones for curvature. At each step, any source with $TS>100$ is fitted with both a power-law and log-parabola spectral shape\footnote{\url{https://fermi.gsfc.nasa.gov/ssc/data/analysis/scitools/source_models.html}}, and the Test Statistic of the curvature ($TS_{\rm CURV}$\footnote{$TS_{\rm CURV} = 2\times(\rm log \mathcal{L}_{\rm LP}-\rm log \mathcal{L}_{\rm PL})$, where $\rm log \mathcal{L}_{\rm LP}$ and $\rm log \mathcal{L}_{\rm PL}$ are, respectively, the log-likelihood of a LP and PL spectrum. The two models are nested, i.e.,\ the PL is a particular case of the LP when $\beta=0$ (see Equations~\ref{eq:pl}-\ref{eq:lp}). Therefore, the value of $TS_{\rm CURV}$ can be used to evaluate the significance of the improvement of the LP spectral model relative to PL.}) is computed. If $TS_{\rm CURV}$ is greater than 16 ($\sim 4\sigma$) then the source is kept as LP; otherwise it is considered as PL. 
Once sources with the highest significance are tested for curvature, we perform a complete fit of the ROI for every iteration of $TS$ and angular separation, freeing all parameters for every source in the field. 
Finally, we delete sources at the edges of the ROI (with an offset to the edge $<1.5\degree$) to minimize the effect of PSF leaking at the ROI edges. Furthermore, to avoid double-counting resulting from the overlap of the ROIs, we choose to retain the source closer to the ROI center if the same source is found in more than one ROI. 
\section{The Real \gm-ray Sky}\label{sec:real}
Our final catalog, obtained by analyzing the actual LAT data through the pipeline described in Section~\ref{sec:pipe}, contains 2680 sources with $TS>25$ (which corresponds to $\sim4\sigma$), at $|b|>20\degree$. 
\begin{table}[th!]
 \begin{center}
	 \caption{Table of total, power-law and log-parabola number of sources in our real detected catalog and the FL8Y with $|b|>20\degree$ and $TS>25$.}\label{tab:real}
	 \begin{tabular}{ c c c}
 \hline
 \hline
	 &  FL8Y & This work \\
 \hline
	 Total & 2930 & 2680 \\
	 Power-Law & 2638 & 2410\\
	 Log-Parabola & 248 & 270\\
 \hline
 \hline
\end{tabular}
\end{center}
\end{table}
To test the consistency of our results with those of FL8Y (which used a different procedure for source detection and optimization), we compare the two catalogs. The FL8Y contains 2930 sources at $|b|>20\degree$ detected from $\rm 100\,MeV$ to $\rm 1\,TeV$ (see Table~\ref{tab:real}). We note that our analysis detects 10\% fewer sources with respect to the FL8Y.
This discrepancy is partially attributed to a thresholding effect, since half of the non-detected FL8Y sources lie close to the detection significance limit ($\sim4\sigma$). In fact, for increasing $TS$ the number of sources in the two catalogs become comparable (e.g., for $TS>36$ the FL8Y contains 2170 sources and our catalog contains 2274). Moreover, we computed the distribution in energy flux of sources in the two catalogs as a function of increasing $TS$. This comparison can be found in Figure~\ref{fig:relnumbers} (right panel). As can be seen, these distributions are very compatible at every TS, with the largest differences towards the lowest fluxes ($\log(S)<-11.3\,\rm erg~cm^{-2}~s^{-1}$). This is expected since faint sources lie at the detection threshold limit. The difference in detection pipelines also plays a role in these discrepancies. 
Nonetheless, the goal of our simplified analysis is not to methodically reproduce the results of the FL8Y, but to produce a stable and reliable detection pipeline that can be consistently used for both real and simulated LAT data sets, allowing us to derive the selection effects of our analysis pipeline.

To further test our detection pipeline, in Figure~\ref{fig:relnumbers} (left panel) we plot the relative numbers of power-law spectra (PL) and log-parabolic (LP) spectra as a function of energy flux in our catalog and the FL8Y. As can be seen, the two distributions are very similar, with a greater fraction of PL spectra at lower fluxes and LPs at higher fluxes\footnote{In the FL8Y there are 44 sources which have a curved spectra modeled by a more complicated shape (power law with exponential cut-off). In our analysis we do not test this shape since the fit would require large computational time to converge, so we exclude these sources from the comparison.}.
Then we cross-match sources positionally in both lists.
To do so, we calculate their angular separations and propagate their 95\% positional errors (i.e., we add them quadratically), evaluated in both our catalog and in the FL8Y (we use the semi-major axis positional error). If the angular separation falls within this 95\% error, the sources are considered to be the same.
We find 2443 positional matches (85\% of the FL8Y).
Considering our choice of positional error and in light of the fact that we detect 10\% fewer sources with respect to the FL8Y, we regard this match as satisfactory. We further check that the number of matches increases for increasing $TS$. Indeed, for $TS>36$ we miss only 10\% of the FL8Y sources and for $TS>49$ only the 5\% induced by the choice of positional error. 
We then use the association listed in the FL8Y to remove from our catalog all sources that are known to be of Galactic origin (i.e.,\ pulsars), since our analysis is not fine tuned for this type of object. Similarly, for the sources significantly spatially extended in the LAT data at $|b|>20\degree$ (Small Magellanic Cloud, Fornax A and Large Magellanic Cloud), we use their position and extension to remove all possible sources that fall within their radius, as we do not perform any test on source extension.
In order to check the consistency of our derived sources' spectral properties with the FL8Y, since the majority of the extragalactic sources detected by the LAT are blazars, we select the matches that are associated with blazars in the FL8Y\footnote{From the associations and identifications in the FL8Y, we select sources listed as BL Lacertae objects (BL Lacs), Flat-Spectrum Radio Quasars (FSRQs) and blazar candidates of uncertain type (BCU).} and have the same spectral type in both catalogs. 
To compare flux values we use the spectral parameters (i.e., pivot energy, flux density and indices) provided by both catalogs to integrate the energy flux between 100\,MeV and 1\,TeV.  
\begin{figure}[t]
\centering
\includegraphics[width=\columnwidth]{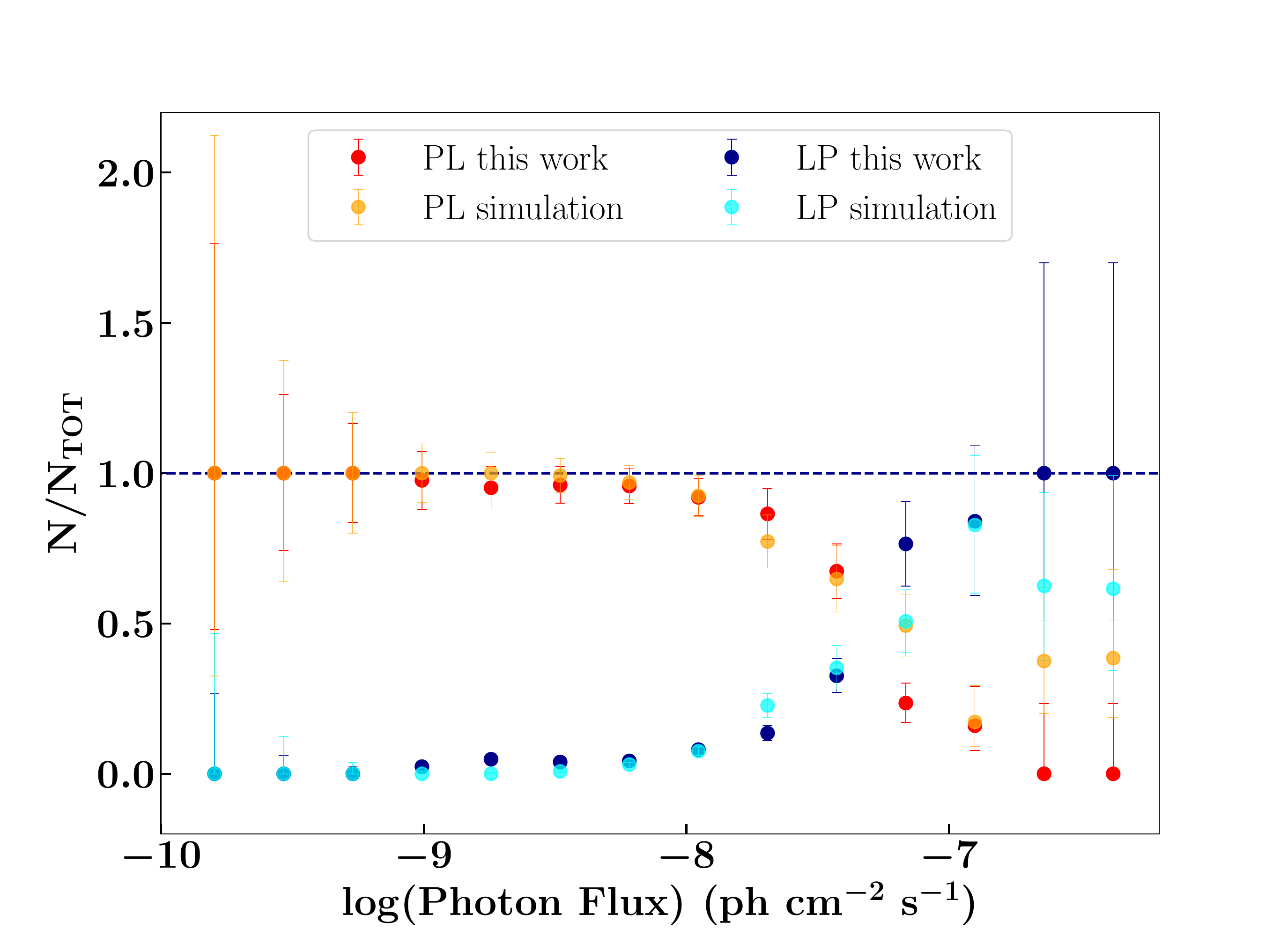}\
\includegraphics[width=\columnwidth]{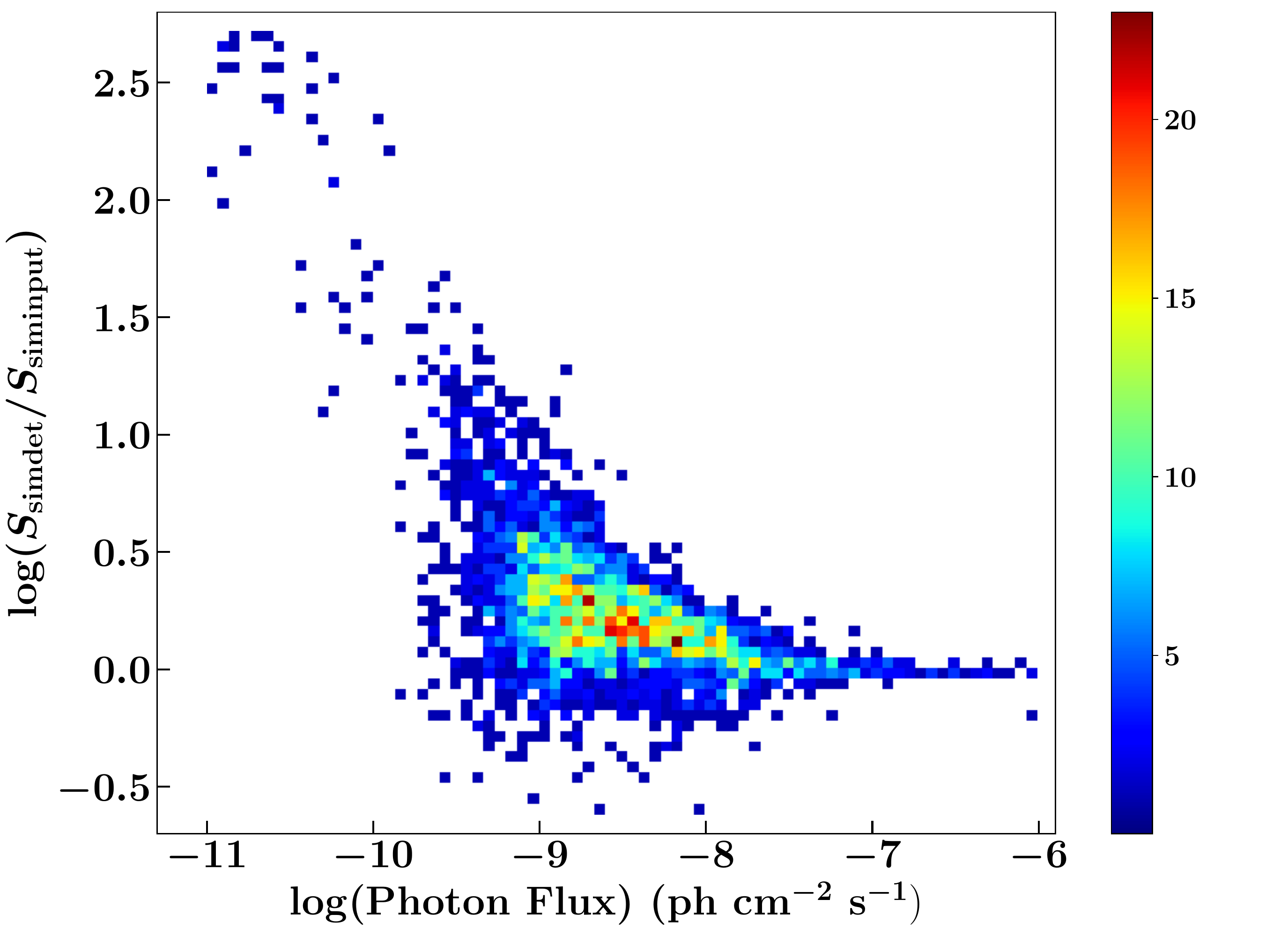}\
	\caption{{\bf Top Panel:} Relative number of power laws and log-parabolas in our real and simulated catalogs as a function of photon flux. The distributions for both spectral types are in good agreement, emphasizing that our simulations represent well the real sky. The larger discrepancy in number of sources at high fluxes is attributed to the low number of sources and large associated errors. {\bf Bottom Panel:} 2d-histogram comparing the photon fluxes ($S$) of sources with positional match between detected simulated and input simulated catalogs (for one simulation). As can be seen, above $\log(S)>-8$\phflux~the flux values are in good agreement. Below this flux, the spread from the one-to-one correlation is larger, although consistent within errors. The apparent overestimate of simulated detected flux below $\log(S)=-9$\phflux~is attributed to the Eddington bias (see Section~\ref{sec:sim}). \label{fig:siminp}}
\end{figure}
In Figure~\ref{fig:ourvs4fgl} we compare the fluxes in the top-left panel for power laws\footnote{Since power-law matches are $>1000$, we chose to plot a 2-d histogram for both flux and indices in order to better understand the distributions.} and in the bottom left panel for log-parabolas. The flux values are in good agreement in both cases, within errors. Indeed, the power-law fluxes are centered along the one-to-one correlation line. The larger spread at lower fluxes is expected and is also associated with larger uncertainties. Considering the statistical errors, all the ratios are consistent with one. There are $\sim10$ outliers with ratios deviating by more than $50\%$ from the one-to-one correlation line, and those are attributed to non-convergent fits\footnote{In our catalog, out of the 2680 sources only $<40$ ($<1.5\%$ of the sample) did not converge properly.}, and hence the extracted source parameters are not reliable. Similarly, the log-parabolas' flux ratio is compatible with the correlation line, taking into account the statistical errors. 
In the top-right panel of Figure~\ref{fig:ourvs4fgl} we compare power-law photon indices. In line with the flux comparison, the indices are in good agreement between the two catalog and their ratio is centered along the correlation line. In the bottom right panel, we show the comparison between photon indices $\alpha_{300}$\footnote{Since the index of a log-parabola depends on the chosen pivot energy, we scale FL8Y indices to our pivot energy, fixed at $300\rm\,MeV$.} and $\beta$ for the log-parabolas. As can be seen, the distributions for $\alpha_{300}$ and $\beta$ are consistent and occupy the same region ($\alpha_{300}>0.5$ and $\beta<0.4$) in the plot for both catalogs. 

\section{The simulated \gm-ray sky} \label{sec:sim}
\begin{table*}[t!]
\scriptsize
\centering 
\caption{Table of input parameters, number of sources and power-law photon indices (mean, $<\Gamma>$, and standard deviation, $\sigma_{\Gamma}$) detected in the real and the simulated sky for $|b|>20\degree$ and $TS>25$. The flux breaks ($S_b$) are listed by their logarithmic value in units of \phflux.}\label{tab:simulated}
 \begin{tabular}{ c c | c c c c c c c | c c c } 
 \hline
	 CATALOG&  \multicolumn{8}{c}{} & N & $<\Gamma>$ & $\sigma_{\Gamma}$  \\
 \hline
 REAL & \multicolumn{8}{c}{} & 2680 & 2.20 & 0.31 \\
	 \hline
	 \hline
      & \multicolumn{1}{c}{Input Shape} & \multicolumn{7}{c}{PARAMETERS} &  &  &  \\
\hline
      & BPL & $\gamma_1$ & $\gamma_2$ &  & & $S_b$   & & & & & \\
      &     & $2.02$     & $1.20$     &  & & $-9.00$   & & & 3258 & 2.21 & 0.28\\
      & DBPL & $\gamma_1$ & $\gamma_2$ & $\gamma_3$ & & $S_{b_1}$ & $S_{b_2}$ & & & & \\
      &      & $1.90$     &  $2.20$    & $1.20$ & & $-8.45$  & $-9.07$  & & 2589 & 2.21 & 0.29 \\
 SIMULATED &  DBPL & $\gamma_1$ & $\gamma_2$ & $\gamma_3$ & & $S_{b_1}$ & $S_{b_2}$ & & & & \\
      &       & $1.90$     &  $2.10$    & $1.10$ & & $-8.40$  & $-9.00$  & &2678 & 2.21 & 0.29 \\
      & TBPL & $\gamma_1$ & $\gamma_2$ & $\gamma_3$ & $\gamma_4$ & $S_{b_1}$ & $S_{b_2}$ & $S_{b_3}$ & & & \\
      &  &  $2.60$  &  $1.60$ &  $2.40$ &  $1.20$ & $-7.22$  & $-8.40$ & $-9.07$ & 3307 & 2.22 & 0.30 \\
	      & DBPL & $\gamma_1$ & $\gamma_2$ & $\gamma_3$ & & $S_{b_1}$ & $S_{b_2}$ & & & & \\
            &      & $1.90$     &  $2.20$    & $1.20$ & & $-8.45$  & $-9.07$ & & 2837 & 2.21 & 0.28\\
 \hline
\end{tabular}
\end{table*}
The LAT survey's biases can be robustly constrained by performing end-to-end Monte Carlo simulations, with the aim of reproducing an extragalactic \gm-ray sky that closely resembles the real one \citep[see e.g.,][]{2010ApJ...720..435A, 2018ApJ...856..106D}. 
We simulate a population of sources randomly distributed in the sky, with spectral characteristics and statistics mimicking the blazar population.
The fluxes are extracted from the range $\rm [10^{-11},10^{-6}]$\phflux, starting an order of magnitude below the minimum detected flux from our catalog. For the very first time, we consider the spectral curvature of the blazars following the recipe detailed
in \citet{2015ApJ...800L..27A}, in order to accurately describe the shape of blazars' \gm-ray spectra between 100\,MeV and 1\,TeV.
We therefore use the following double power-law input shape:
\begin{equation}
	\frac{dN}{dE}=K\left[\left(\frac{E}{E_b}\right)^{\delta_1}+\left(\frac{E}{E_b}\right)^{\delta_2}\right]^{-1} 
\label{eq:spec}
\end{equation}
where $E_b$ is the break energy calculated from the $E_b-\Gamma$ correlation found in \citet{2015ApJ...800L..27A}, and $\Gamma$ is the power-law photon index of a blazar's \gm-ray spectrum. First, a $\Gamma$ is randomly drawn from a Gaussian distribution of mean 2.45 and standard deviation of 0.40. This is transformed into $E_b$ 
as $\log E_{b}=9.25-4.11\Gamma$ \citep[following][]{2015ApJ...800L..27A}. The indices of Equation~\ref{eq:spec} are set to be $\delta_1=1.7$ and $\delta_2=2.8$. For the latter we use the value reported in \citealp[][]{2018ApJ...856..106D} to reproduce the source-count distribution of the Third Catalog of Hard LAT Sources (3FHL, \citealp[][]{2017ApJS..232...18A}), which contains sources detected by the LAT above 10 GeV. 
\begin{figure*}[ht!]
\includegraphics[width=0.33\textwidth]{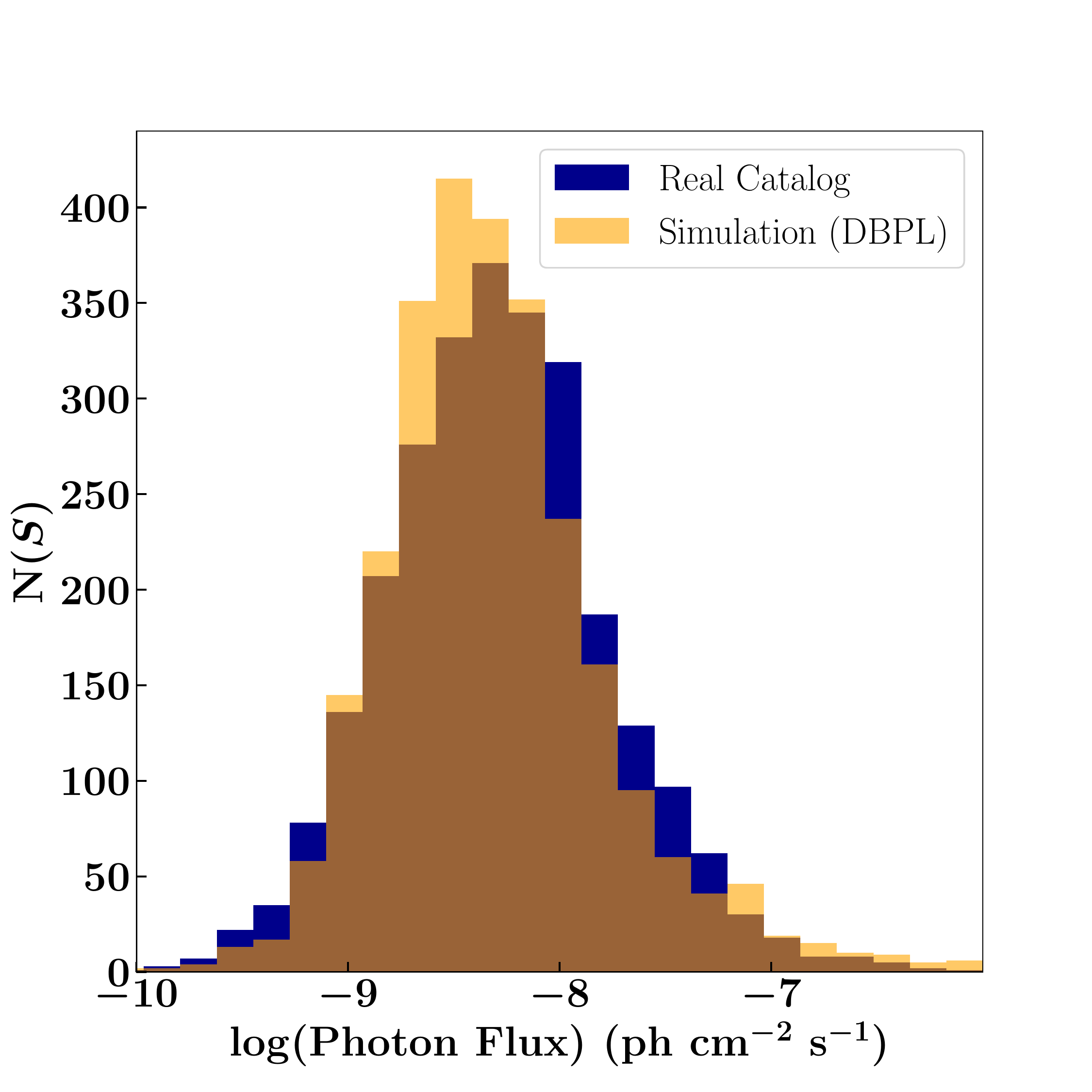}
\includegraphics[width=0.33\textwidth]{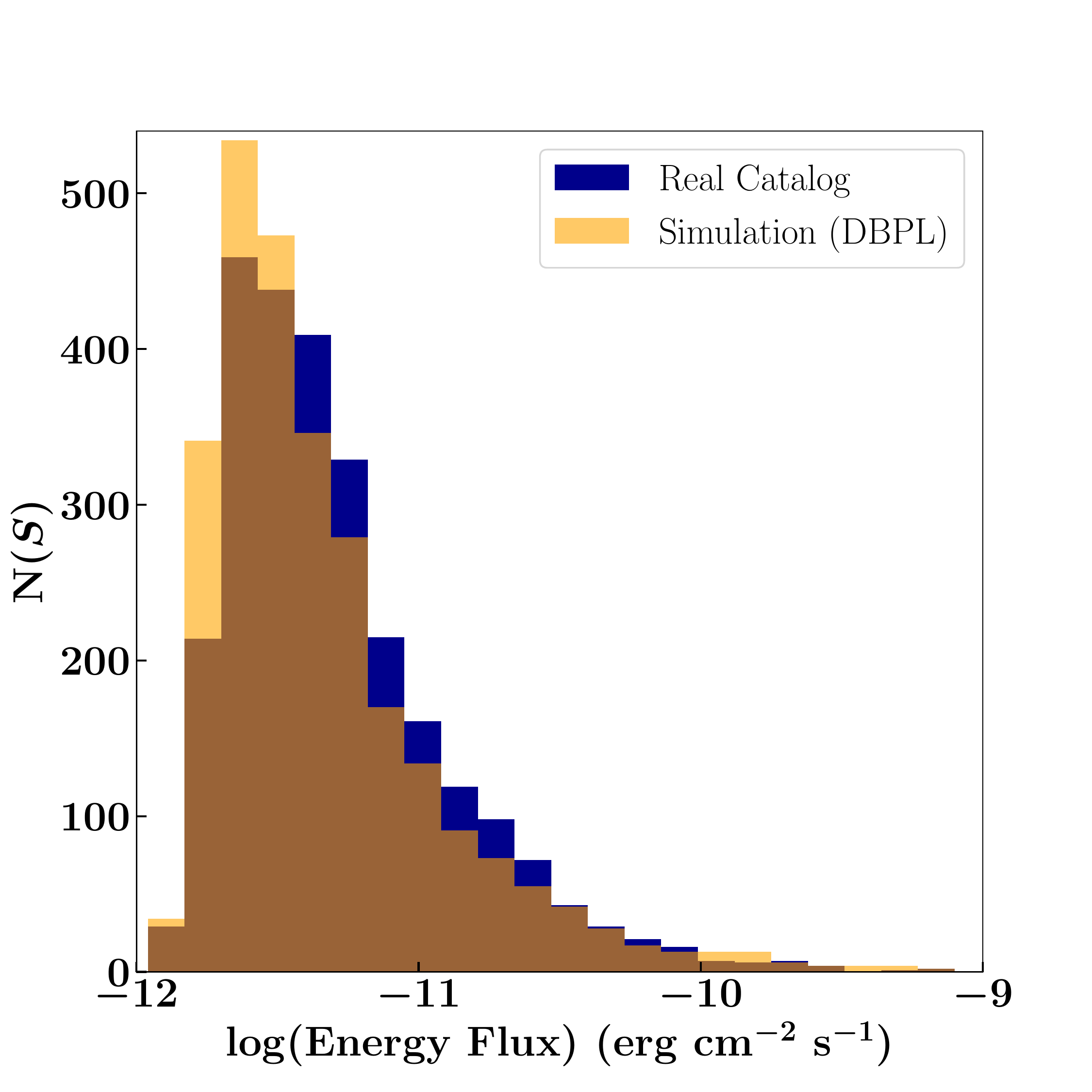}
\includegraphics[width=0.33\textwidth]{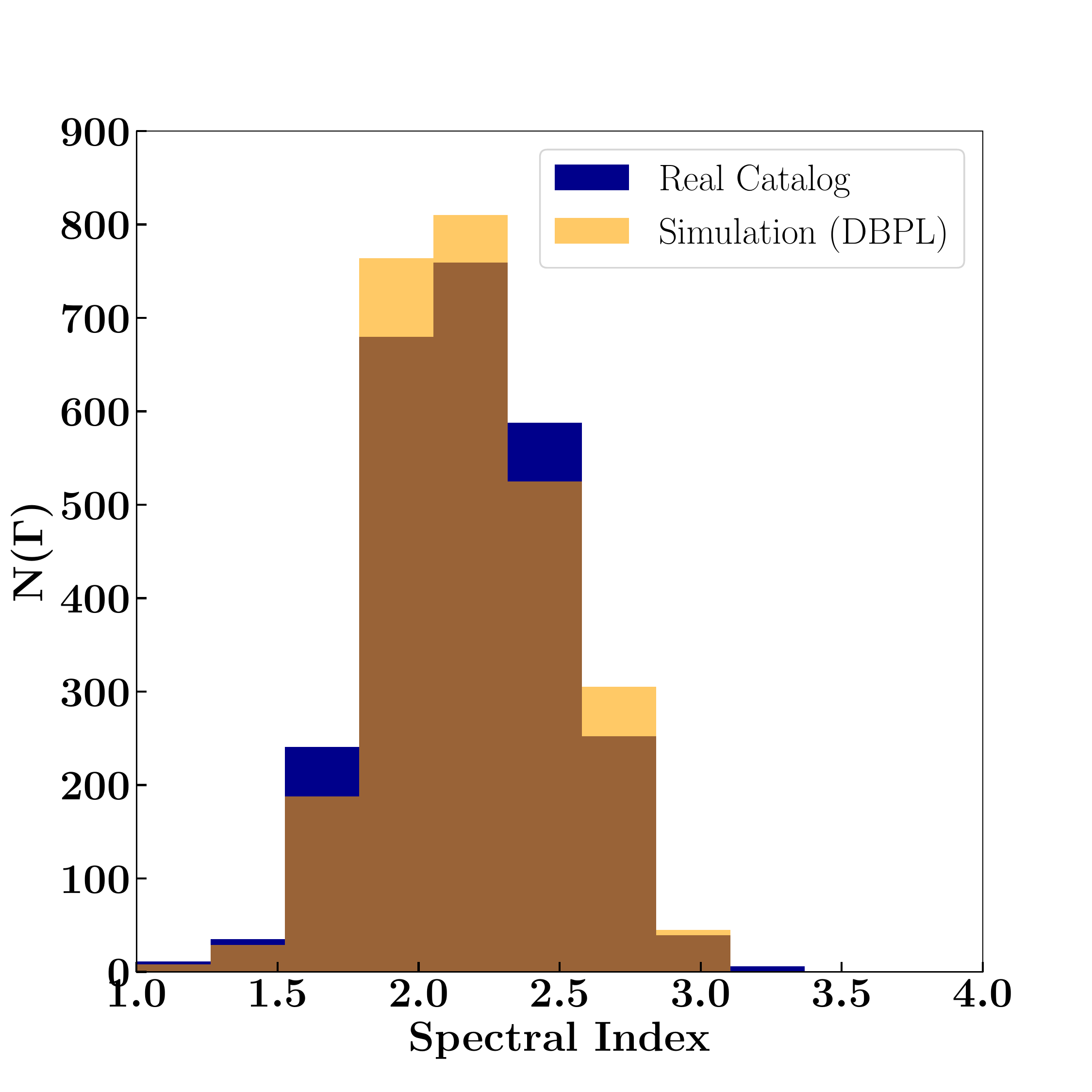}
\caption{{\bf From left to right}: Histogram of detected photon flux, detected energy flux and detected spectral indices for one simulation (DBPL), and the real catalog. As can be seen, all simulated distributions are consistent with the real ones, implying that the simulations are a good representation of the real sky. All other simulations have similar distributions.\label{fig:distr_sim}}
\end{figure*}

In order to produce Monte Carlo simulations that accurately represent the true $\gamma$-ray sky, knowledge of the intrinsic {\it logN-logS} is needed. However, this is not known a priori, but results from this work. To cope with this, the input photon flux {\it logN-logS} in the simulations was changed until it matched the reconstructed one reasonably well.
The input shapes used in this analysis for the differential {\it logN-logS} ($dN/dS$) are the following: 
\begin{enumerate}
	\item Broken power law (BPL): 
\begin{equation}\label{eq:2}
 \frac{dN}{dS} = K \begin{cases}
   S^{-\gamma_1} &  S>S_b\\
    S^{-\gamma_2}S_b^{-\gamma_1+\gamma_2} & S\leq S_b
 \end{cases}
\end{equation}
where $\gamma_1$ and $\gamma_2$ are the slopes after and before the break $S_b$, respectively.
\item Double broken power law (DBPL): 
\begin{equation}\label{eq:3}
\resizebox{.9\hsize}{!}{
 $\frac{dN}{dS} = K \begin{cases}
     S^{-\gamma_1} &  S>S_{b1}\\
     S^{-\gamma_2}S_{b_1}^{-\gamma_1+\gamma_2} & S_{b_2}< S \leq S_{b_1}\\
     S^{-\gamma_3}S_{b_1}^{-\gamma_1+\gamma_2} S_{b_2}^{-\gamma_2+\gamma_3} & S\leq S_{b_2}
 \end{cases}
        $}
\end{equation}
where $\gamma_1$ is the slope of the distribution before the first break $S_{b_1}$, $\gamma_2$ the slope between the first and second break ($S_{b_2}$), and $\gamma_3$ is the slope after $S_{b_2}$.
\item Triple broken power law (TBPL): 
\begin{equation}\label{eq:4}
\resizebox{.9\hsize}{!}{
 $\frac{dN}{dS} = K \begin{cases}
     S^{-\gamma_1} &  S>S_{b_1}\\
     S^{-\gamma_2}S_{b_1}^{-\gamma_1+\gamma_2} & S_{b_2}<S \leq S_{b_1}\\
     S^{-\gamma_3}S_{b_1}^{-\gamma_1+\gamma_2} S_{b_2}^{-\gamma_2+\gamma_3} & S_{b_3}< S\leq S_{b_2}\\
     S^{-\gamma_4}S_{b_1}^{-\gamma_1+\gamma_2} S_{b_2}^{-\gamma_2+\gamma_3} S_{b_3}^{-\gamma_3+\gamma_4} & S\leq S_{b_3}\\
 \end{cases}
        $}
\end{equation}
where $\gamma_1$ is the slope of the distribution before the first break $S_{b_1}$, $\gamma_2$ the slope between the first and second break ($S_{b_2}$), $\gamma_3$ the slope between the second and third break ($S_{b_3}$) and $\gamma_4$ the slope after $S_{b_3}$.
\end{enumerate}
\begin{figure*}[t!]
\includegraphics[width=0.47\textwidth]{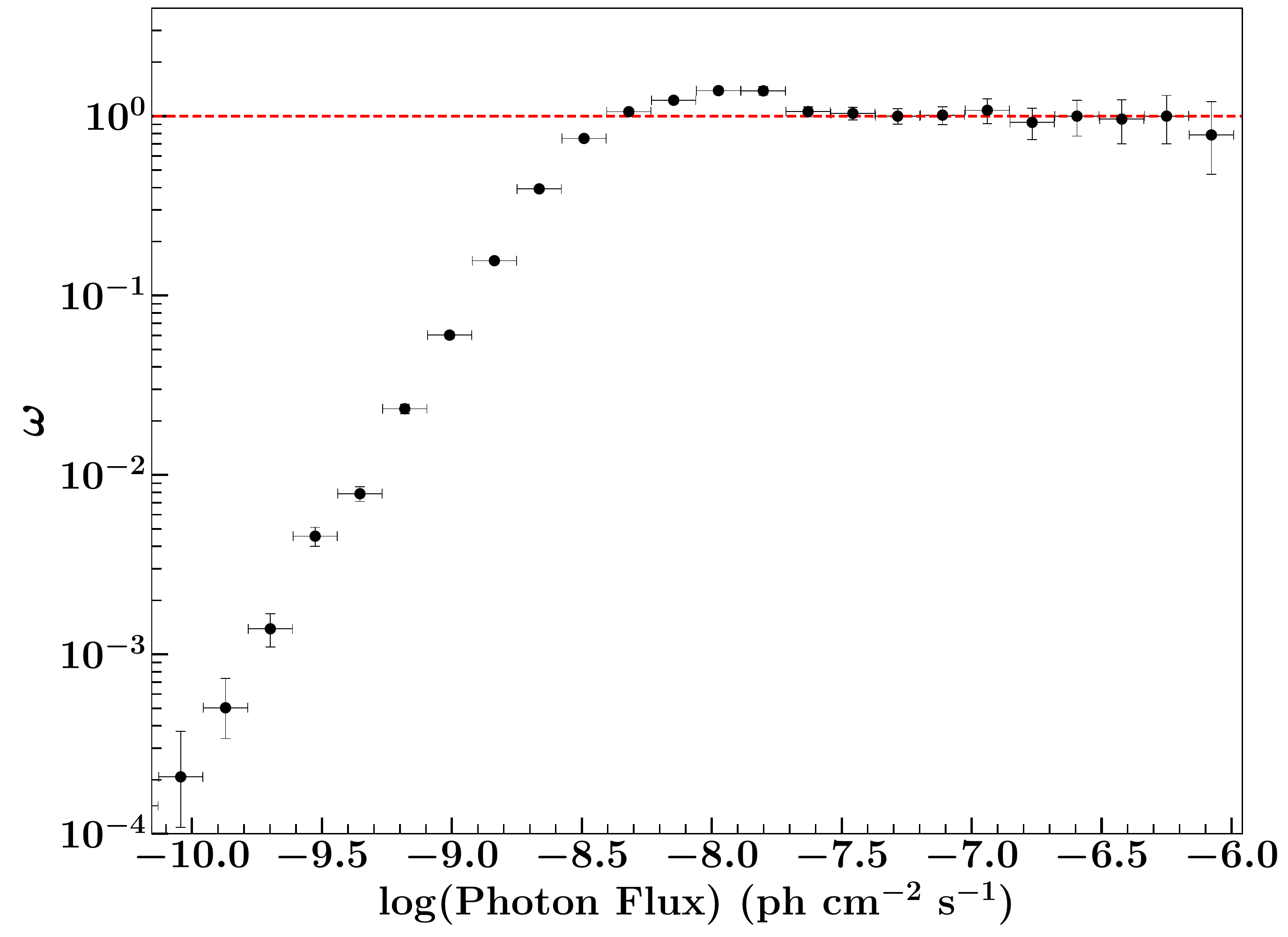}\
\includegraphics[width=0.53\textwidth]{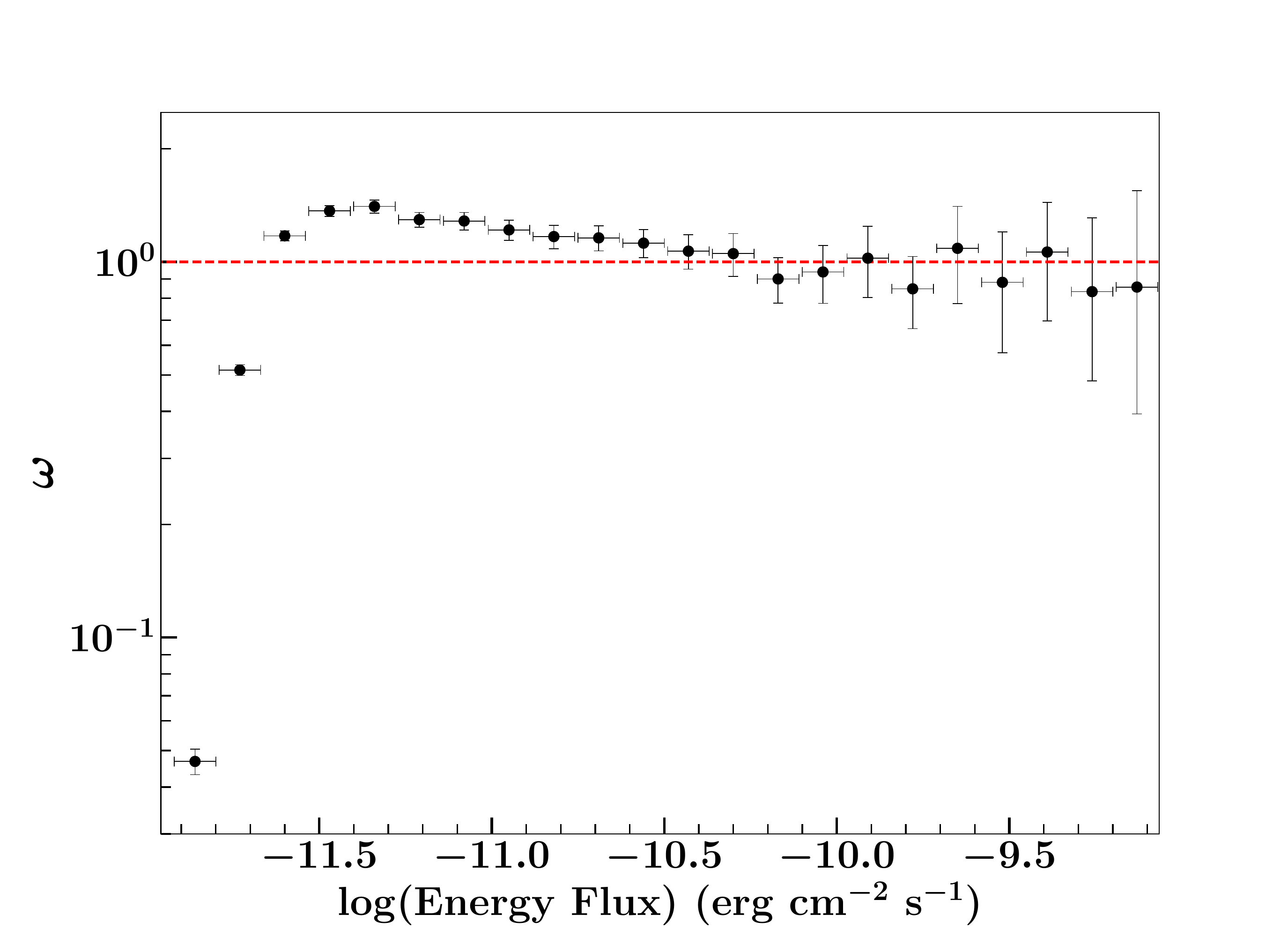}           
	  \caption{Efficiency ($\omega$) plotted as a function of measured photon flux (left panel) and energy flux (right panel) obtained with four simulations. Both functions follow a similar trend: above $\log(S)=-7.7$\phflux ($-10.6$\ergflux) the efficiency is equal to one, corresponding to a perfect detection capability of the LAT. In the range $\log(S)=[-8,-7.7]$\phflux ($[-11.6,-10.6]$\ergflux), the efficiency is greater than one, implying that the LAT detects more sources at these fluxes than actually present (Eddington bias). The efficiency as a function of photon flux has a slower decrease than the efficiency as a function of energy flux; this is due to the strong dependence of the photon flux on the spectral index. The error bars are computed in both cases using Poisson statistics.\label{fig:eff}}
\end{figure*}

For all of the above, $K$ is the appropriate normalization constant.
We generate a total of five simulations, where four use the standard IEM (one BPL, two DBPL and one TBPL), and one uses the alternative IEM (DBPL; this is needed to evaluate the systematics of the analysis, see Section~\ref{sec:syst}).
The values for the parameters employed in the various simulated input catalogs can be found in Table~\ref{tab:simulated}.
Each list of synthetic sources is used to 
generate a simulation of the sky in each ROI (using the tool \texttt{simulate\_roi}). Then the pipeline described in Section~\ref{sec:pipe} is applied to blindly detect sources in the simulated sky. The normalization of the isotropic template is set to 0.7 for all simulations.
In Table~\ref{tab:simulated}, we report the number of sources detected and the mean power-law photon index for the simulated catalogs and the real sky. The parameters in all simulations are consistent with the real ones. In the top panel of Figure~\ref{fig:siminp}, we show the ratios of power-law and log-parabola sources relative to the total, both for the real sky and for one simulation. 
As can be seen, the two distributions are in good agreement: the fraction of power-law spectra is higher at lower fluxes and decreases towards higher ones, while the log-parabolic spectra follow the opposite trend. Furthermore, for every simulation we checked the distribution of sources as a function of detected photon flux, energy flux and spectral index. In Figure~\ref{fig:distr_sim} we show these comparisons for one simulation. All simulations are in agreement with the distributions found for our real catalog, reflecting the close resemblance of the simulated sky to the real one. 
The bottom panel of Figure~\ref{fig:siminp} shows the flux ratio of the simulated detected sources to the input ones. The comparison has been done for positionally matching sources, i.e.,\ with angular distance within the 99 \% positional error. As can be seen, at bright fluxes ($\log(S)>-8$\phflux) both populations follow the one-to-one correlation. At lower fluxes the spread of these ratios is larger, though still lying on the correlation line if considering the statistical errors. Below $\log(S)\sim-9$\phflux, it instead appears that detected simulated sources have significantly brighter flux value with respect to the simulated input ones.
\begin{figure*}[t!]
	\centering
\includegraphics[width=0.48\textwidth]{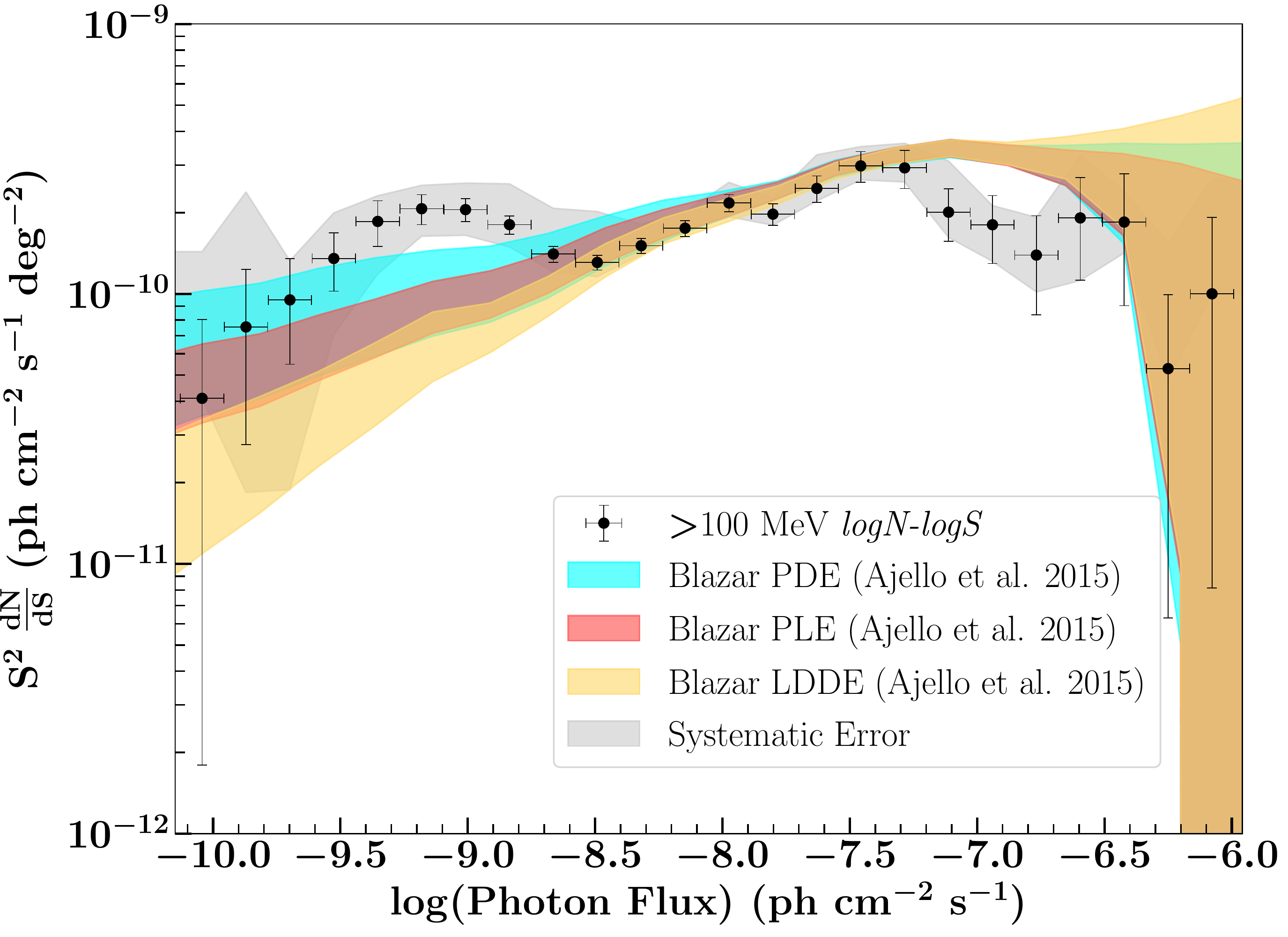}\
\includegraphics[width=0.48\textwidth]{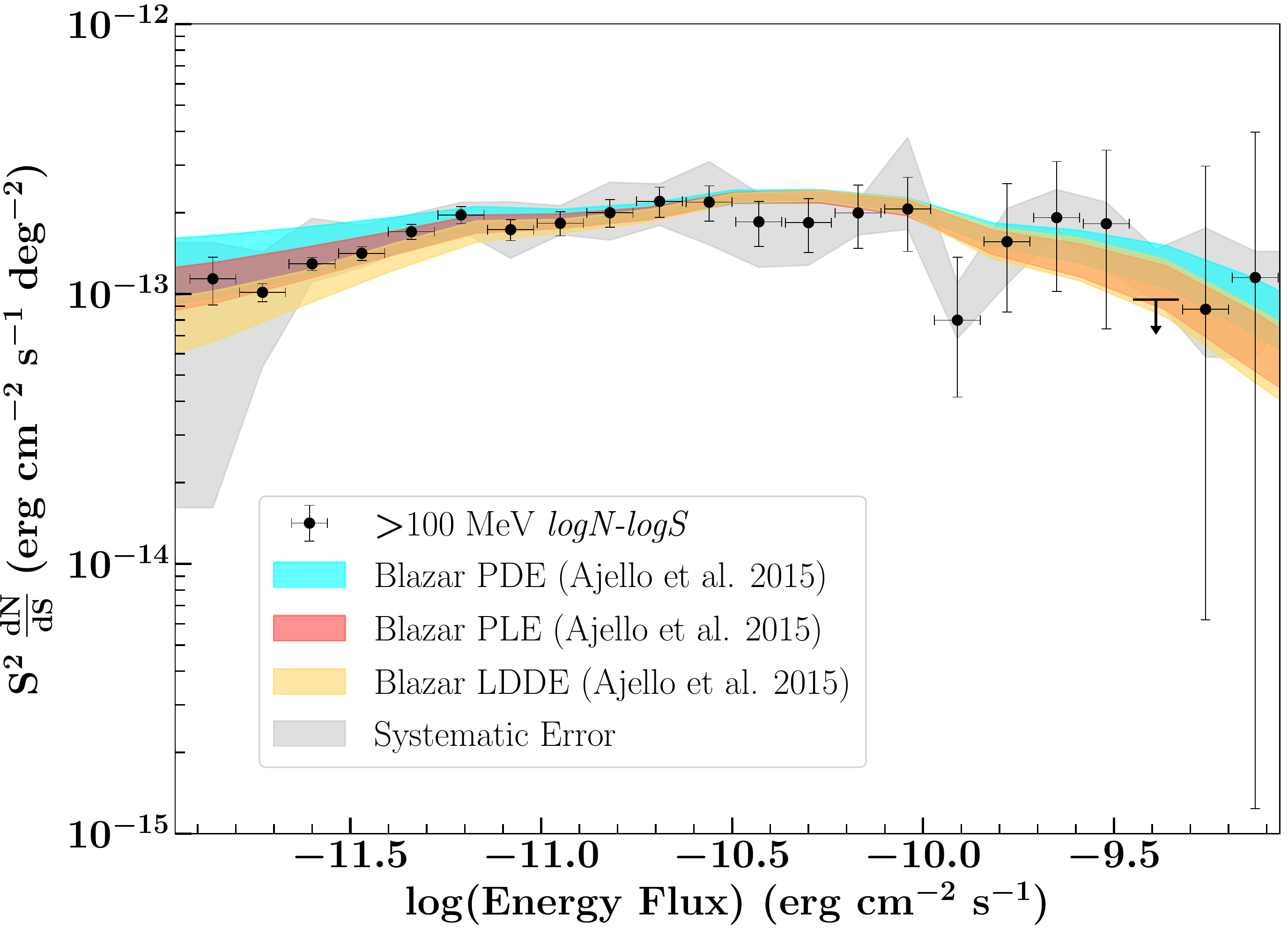}\
\includegraphics[width=0.48\textwidth]{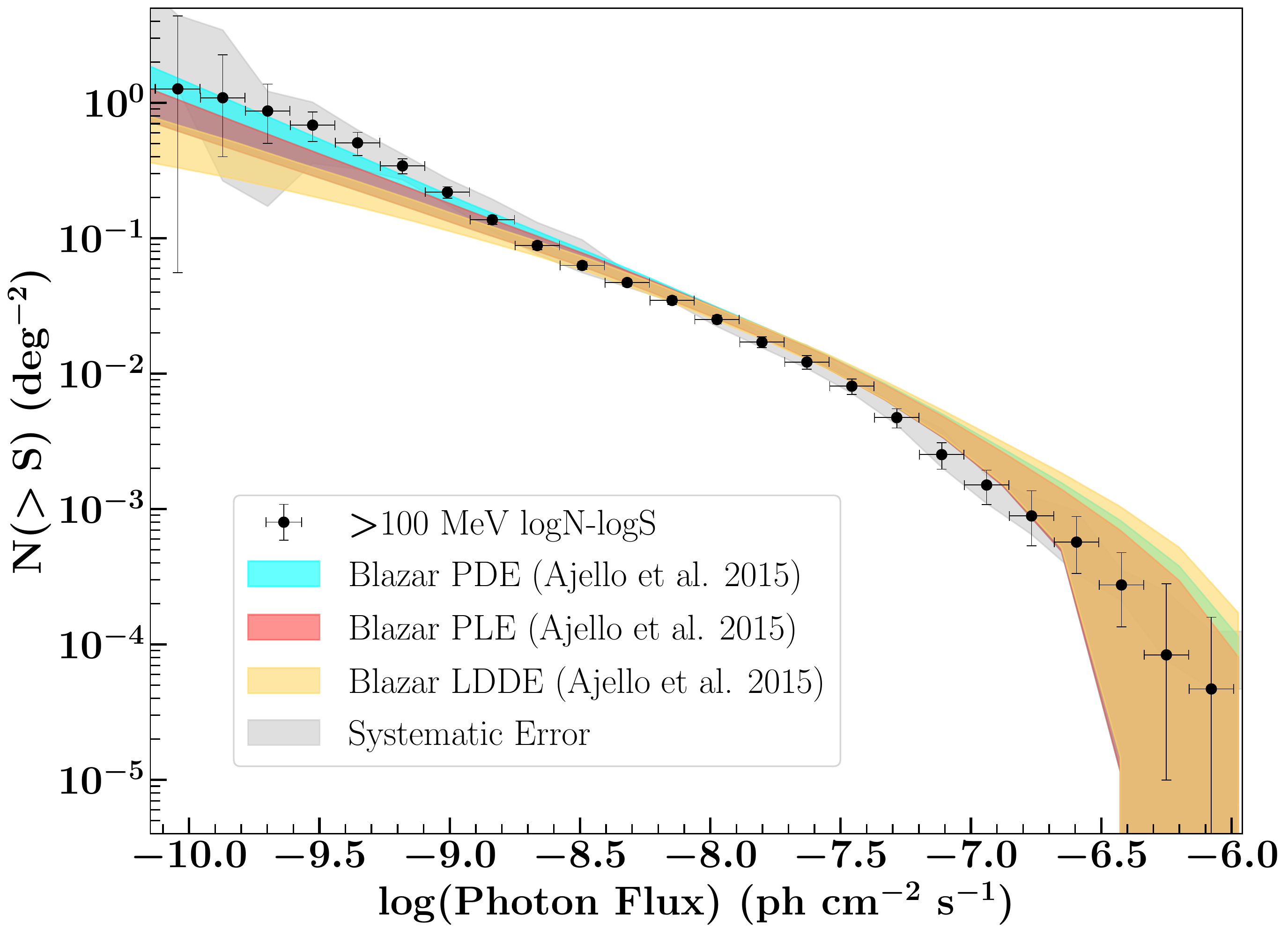}\
\includegraphics[width=0.48\textwidth]{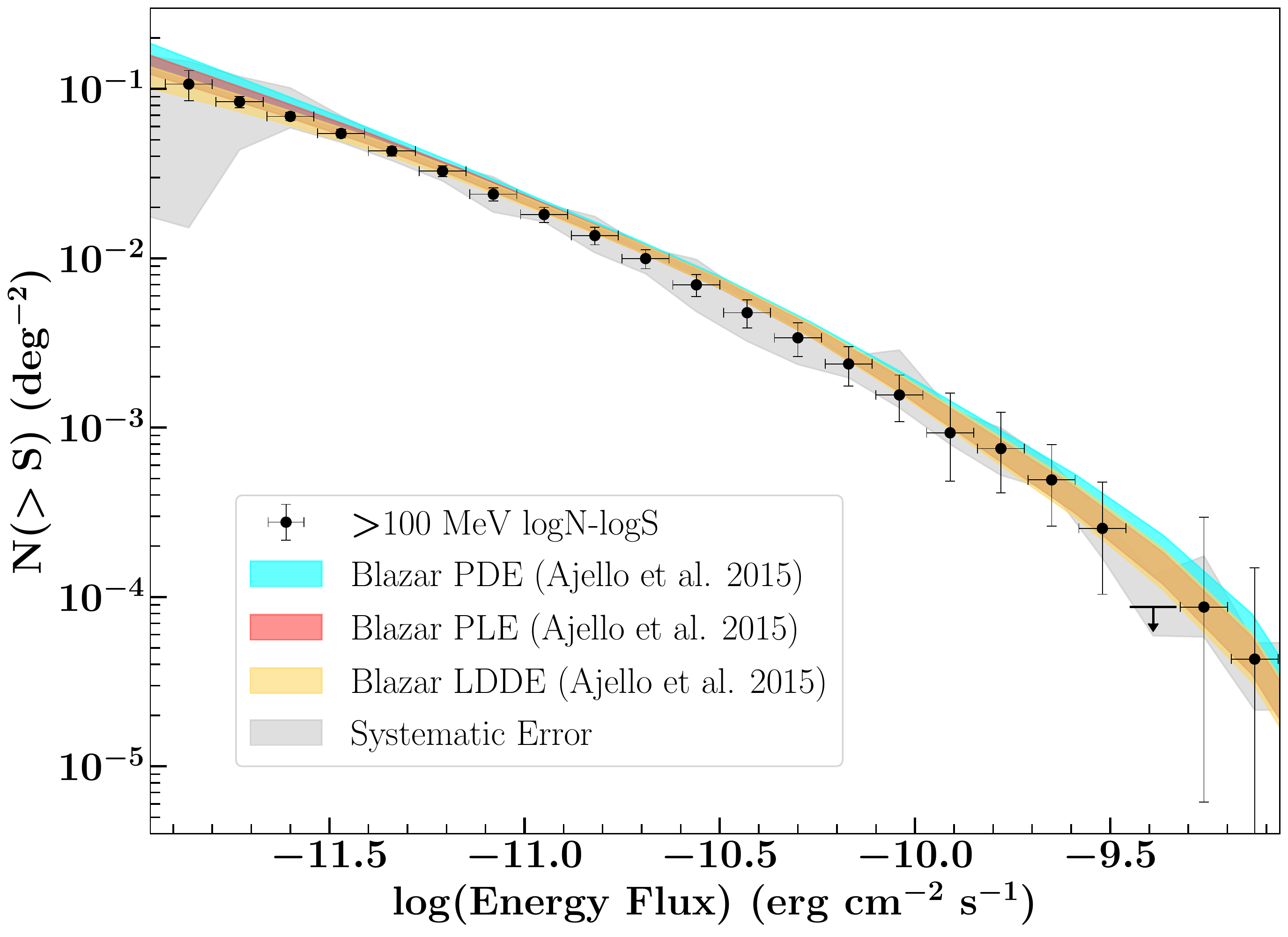}\
	
	  \caption{Differential (top) and cumulative (bottom) intrinsic source-count distribution ({\it logN-logS}, black data points) of point sources plotted as a function of measured photon flux (left panel) and energy flux (right panel). The gray shaded region represents the systematic errors. The cyan, red and orange shaded bands represent, respectively, the Pure Density Evolution (PDE), the Pure Luminosity Evolution (PLE) and the Luminosity-Dependent Density Evolution (LDDE) model predictions from \citet{2015ApJ...800L..27A}. The PDE model is the one that best represents our {\it logN-logS}. We underline how our analysis, using 8 years of the Pass 8 data, has reached $\sim10^{-10}$\phflux ($\sim10^{-12}$\ergflux), an order of magnitude lower than \citet{2010ApJ...720..435A}. The energy flux {\it logN-logS} is relatively flatter than the photon flux one, due to the low dependence of the energy flux on the source spectral shape. \label{fig:lognlogs}}
\end{figure*}
This effect is identified as the Eddington bias \citep{1913MNRAS..73..359E}. The flux ($F$) from astrophysical sources has a fluctuation of $\pm \Delta F$. If a source falls closely to the detection threshold of the instrument, it would be more easily detected if $F'=F+\Delta F$. Therefore, such objects ($\sim 3\%$ of our sample) are found with a higher flux than their intrinsic one. Input to the simulations do not suffer this bias. Hence
the flux ratio will reflect this overestimate of detected flux. We also check for the presence of spurious sources in every simulation using the method employed by \citet{2018ApJ...856..106D}, and we find that they are $<2\%$ in all samples.

\section{Detection Efficiency and Intrinsic Source Count Distribution}\label{sec:eff}
Once the simulated skies have been analyzed with the same pipeline as the real one, one can calculate the detection efficiency ($\omega$) of the LAT high-latitude survey as:
\begin{equation}
	\omega(S_i)=\frac{N_{\rm simdet}(S_i)}{N_{\rm siminput}(S_i)}
        \label{eq:eff}
\end{equation}
where $S_i$ is the photon (or energy) flux in the $i$th bin, $N_{\rm simdet}(S_i)$ is the number of (simulated) sources detected with a (measured) flux $S_i$ in all simulations, and $N_{\rm siminput}(S_i)$ is the number of sources simulated with the same flux (in all simulations). 
Traditionally, the efficiency of the LAT is presented as a function of photon flux. Due to the fact that the photon flux highly depends on the spectral index with which the sources are modeled and detected (e.g.,\ sources with harder spectral index will be detected to lower flux values than the softer ones; \citealp[see][]{2010ApJ...720..435A}), here we present the efficiency (and later the {\it logN-logS}) obtained with both photon and energy flux. The latter has a lower dependence on the sources' spectral indices, and therefore produces more reliable and stable outcomes.
The results are shown in Figure~\ref{fig:eff},
where the left panel shows the efficiency as a function of photon flux and the right panel shows the efficiency, as a function of energy flux. As expected, for flux values higher than $\log(S)>-7.7$ \phflux ($>-10.6$ \ergflux), $\omega(S)$ is 1, i.e.,\ the probability of LAT detecting sources at high flux values is 100\%. Around $\log(S)\sim-8$ \phflux ($-11$ \ergflux) the function rises slightly above 1, as a consequence of the Eddington bias. Afterwards the efficiency slowly drops to zero. We note that this decrease is sharper for the energy flux case, since this quantity is less sensitive to the sources' spectral shape and hence its flux variations. The errors are evaluated using Poisson statistics.
\begin{figure}[t]
	\includegraphics[width=\columnwidth]{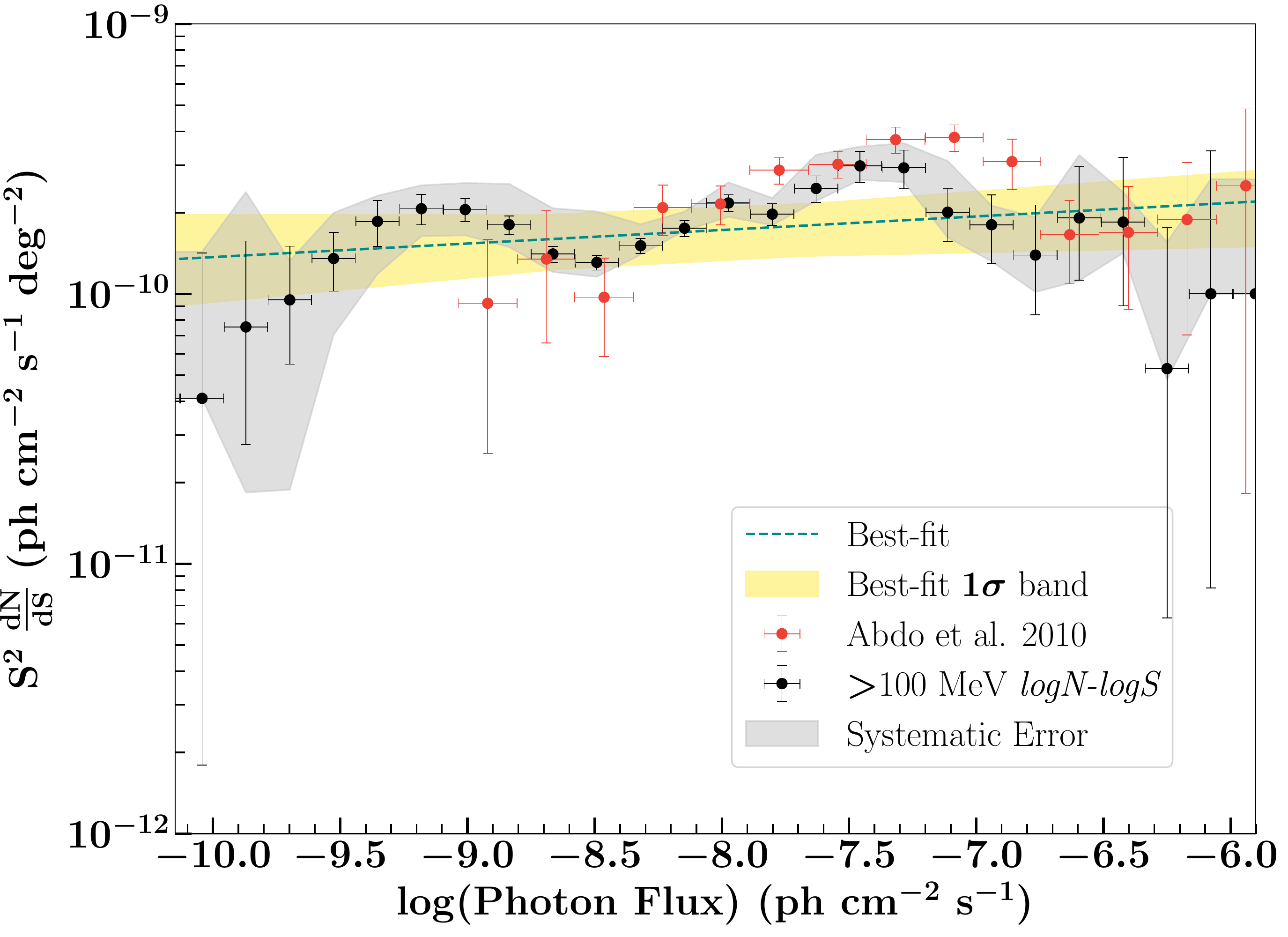}
	\includegraphics[width=\columnwidth]{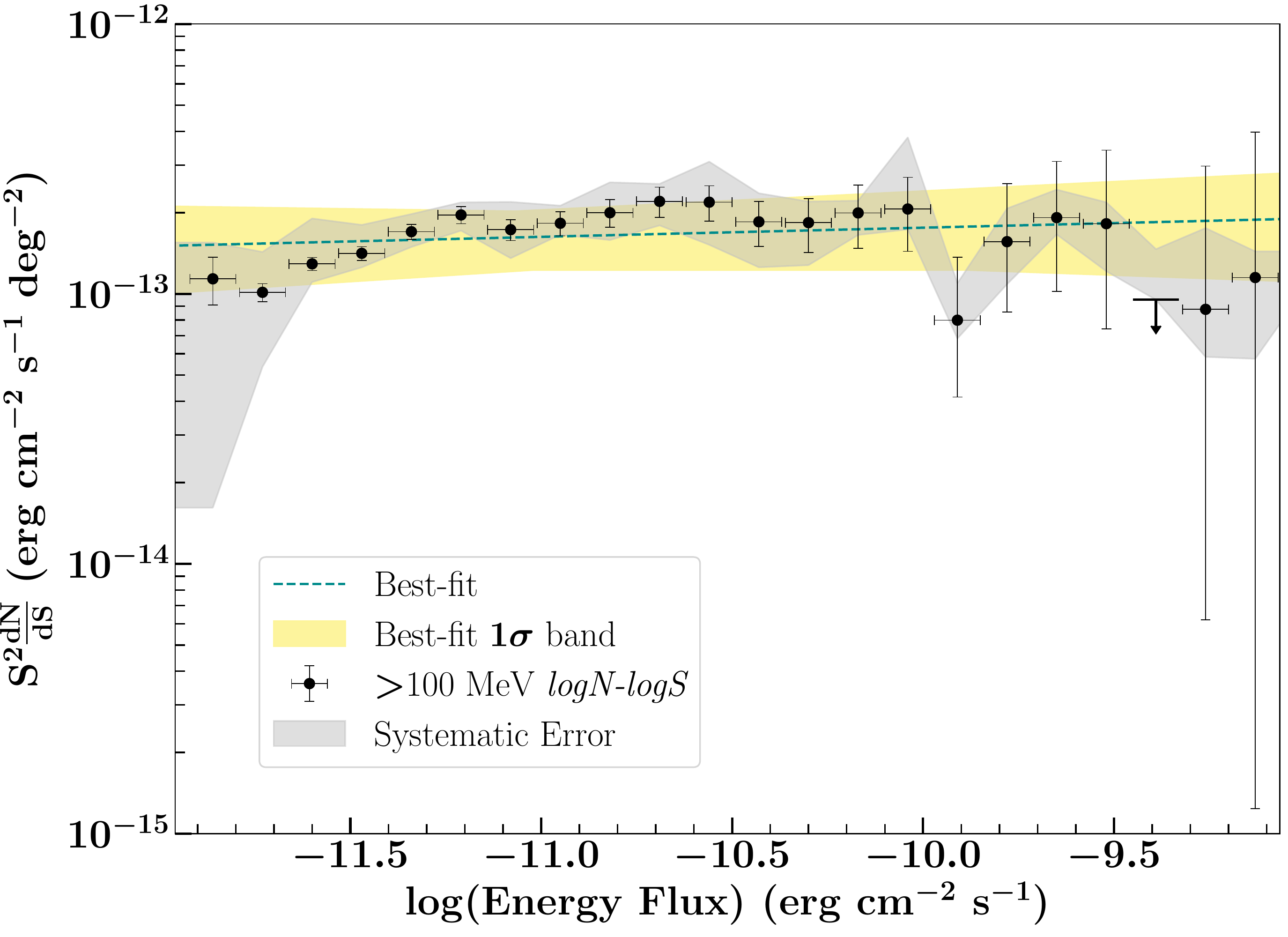}
	\caption{Best-fit shapes obtained for both the photon (top) and energy (bottom) flux {\it logN-logS}. The fit for both the photon and energy flux {\it logN-logS} favors a PL shape. For comparison, we have inserted in the top figure the {\it logN-logS} derived in \citet{2010ApJ...720..435A}, which falls within the 1$\sigma$ error band derived from our best-fit, corroborating the analysis.\label{fig:best-fit}}
\end{figure}

Finally, we can correct our real catalog of point sources with $\omega(S)$ in order to obtain their intrinsic flux distribution and to understand their contribution to the EGB.
The differential {\it logN-logS} is given by:
\begin{equation}
	\frac{dN}{dS}=\frac{1}{\Omega \Delta S_i}\frac{N_i}{\omega(S_i)}
\end{equation}
where $\Delta S_i$ is the width of the flux bin centered at $S_i$, $N_i$ is the number of sources detected in that flux bin, and $\Omega$ is the solid angle of the sky at $|b|>20\degree$.
In Figure~\ref{fig:lognlogs}, the {\it logN-logS} is shown in its differential (top) and cumulative (bottom) forms as a function of photon and energy flux. 
For the first time, the {\it logN-logS} is characterized down to $\sim$$10^{-10}$\phflux ($\sim$$10^{-12}$\ergflux), an order of magnitude lower than that achieved by \citet{2010ApJ...720..435A}, making this the deepest $\gamma$-ray {\it logN-logS} to date. 
As will be demonstrated later in Section~\ref{sec:complete}, the population of point sources underlying this distribution are mainly blazars. 

In order to determine the spectral shape that best represents the intrinsic population of point sources, we perform statistical fits using the summation in quadrature of statistical and systematic (Section~\ref{sec:syst}) errors for both photon flux and energy flux cases.
Along with a BPL, DBPL and TBPL, we test the following shapes
\begin{enumerate}
	\item Power law (PL):
		\begin{equation}
			 \frac{dN}{dS} = K S^{-\gamma}
                \label{eq:pl} 
		\end{equation}
              where $\gamma$ is the slope of the distribution.
	\item  Log-Parabola (LP):
		\begin{equation}
			\frac{dN}{dS}=K\left(\frac{S}{S_0}\right)^{-\alpha+\beta \log(\frac{S}{S_0})}
                \label{eq:lp}
                \end{equation}
               where $\alpha$ is the slope of the distribution and $\beta$ its curvature, or the slope of the distribution at $S_0$ pivot energy.
\end{enumerate}
For all the above spectral models, $K$ is the appropriate normalization constant. When the fit results in a reduced $\chi^2$ close to one, and the improvement on $\Delta\chi^2>3\sigma$, we consider the shape a good representation of our distribution. We find that for the both the photon flux and energy flux {\it logN-logS}, the PL is the best representation for the intrinsic {\it logN-logS}, with a reduced $\chi^2_{\rm }$ of 1.28 and 0.70, respectively. The best-fit indices are: for the photon flux {\it logN-logS} $\gamma=1.94\pm0.02$; for the energy flux {\it logN-logS} $\gamma=1.96\pm0.04$. This also corroborates the fact that at low flux values the distribution remains flat.
Figure~\ref{fig:best-fit} shows the best-fit shapes for both photon and energy flux.

\subsection{Maximum-Likelihood Fit}\label{sec:mlfit}
To check the soundness of our result, we further employ a maximum likelihood (ML) fit 
which follows the methodology detailed in \citet{2010ApJ...720..435A}.
In our case, we adopt the normalization-free form of the likelihood function defined in \citet{2006ApJ...643...81N} which can be written as
\begin{equation}
\mathcal{L} = \prod_{i=0}^{N_{\rm obs}} \frac{1}{N_{\rm exp}}\phi(S_i)
\end{equation}
where $N_{\rm obs}$ is the total number of sources and $S_i$ is the photon (or energy) flux of the $i$th source; $\phi(S)$ is defined as
\begin{equation}
\phi(S) = \frac{dN}{dS}\omega(S)
\end{equation}
where $dN/dS$ is one of the tested spectral forms (Equations~\ref{eq:2}-\ref{eq:4} and \ref{eq:pl}-\ref{eq:lp}); $N_{\rm exp}$ is the expected number of sources and can be evaluated as:
\begin{equation}
N_{\rm exp} = \int_{S_{\rm min}}^{S_{\rm max}} \phi(S) dS 
\end{equation}
where $S_{\rm min}=10^{-11}\phflux$ and $S_{\rm max}=10^{-6}\,{\rm ph~cm^{-2}}\\
{\rm s^{-1}}$ for the photon flux case, and $S_{\rm min}=10^{-13}\ergflux$ and $S_{\rm max}=10^{-9}\,{\rm ph~cm^{-2}}
{\rm s^{-1}}$ for the energy flux case.

The standard $C=-2\ln(\mathcal{L})$ is then calculated as:
\begin{equation}
C = -2\left[\left(\sum_{i=0}^{N_{\rm obs}}\ln\phi(S_i)\right) - N_{\rm obs}\ln\left(N_{\rm exp}\right)\right]
\end{equation}
The best-fit parameters and their associated $1\sigma$ errors are computed by varying the
parameters of interest and minimizing the value of $C$ until an improvement of $\Delta C=1$ is achieved (under the assumption that $\mathcal{L}\propto\exp(-\chi^2/2)$, see e.g.~\citealp{1989NYASA.571..601L,2006ApJ...643...81N}). For this purpose we use 
the \texttt{pyROOT} implementation of \texttt{Minuit}\footnote{\url{https://root.cern.ch/doc/master/classTMinuit.html}}. 
Once the $C$ values are extracted for all models, the Bayesian Information Criteria (BIC, \citealp{bayesfactor}) is employed to determine which model provides the best-fit to the data, i.e.~the model with the lowest BIC. The different BIC values are listed in Table~\ref{tab:BIC}. In our results we point out the following: 1)~although the BIC values for the curved models in the energy flux case
are slightly lower than that for the PL shape,
within the $1\sigma$ errors the parameter values for both LP and BPL are consistent with the PL case, and 2)~for the photon flux {\it logN-logS} a difference of 2 in BIC values between PL and TBPL models represents a negligible
improvement of the fit and cannot be regarded as significant. 
Therefore, in agreement with the reduced $\chi^2$
results, the best-fit spectral shape for both the photon and energy flux distributions is the power law.
The best-fit indices are: $\gamma=1.90\pm0.01$ for the photon flux and $\gamma=1.92\pm0.02$ for the energy flux, 
consistent with the results of the previous section. 

\begin{table}[b!]
\centering
\caption{BIC values~\label{tab:BIC} derived from the ML fits (Section~\ref{sec:mlfit})}
\begin{tabular}{ c | c c }
 & Photon Flux & Energy Flux \\
\hline
PL   & -93539.1  & -133091.1 \\
BPL  & -93533.7  & -133097.3 \\
LP   & -93538.01 & -133097.3 \\
DBPL & -93537.3  &    - \\
TBPL & -93541.8  &    - \\
\end{tabular}
\end{table}
\subsection{LogN-logS Systematics}\label{sec:syst}
\begin{figure}[t]
\centering
\includegraphics[width=1.\columnwidth]{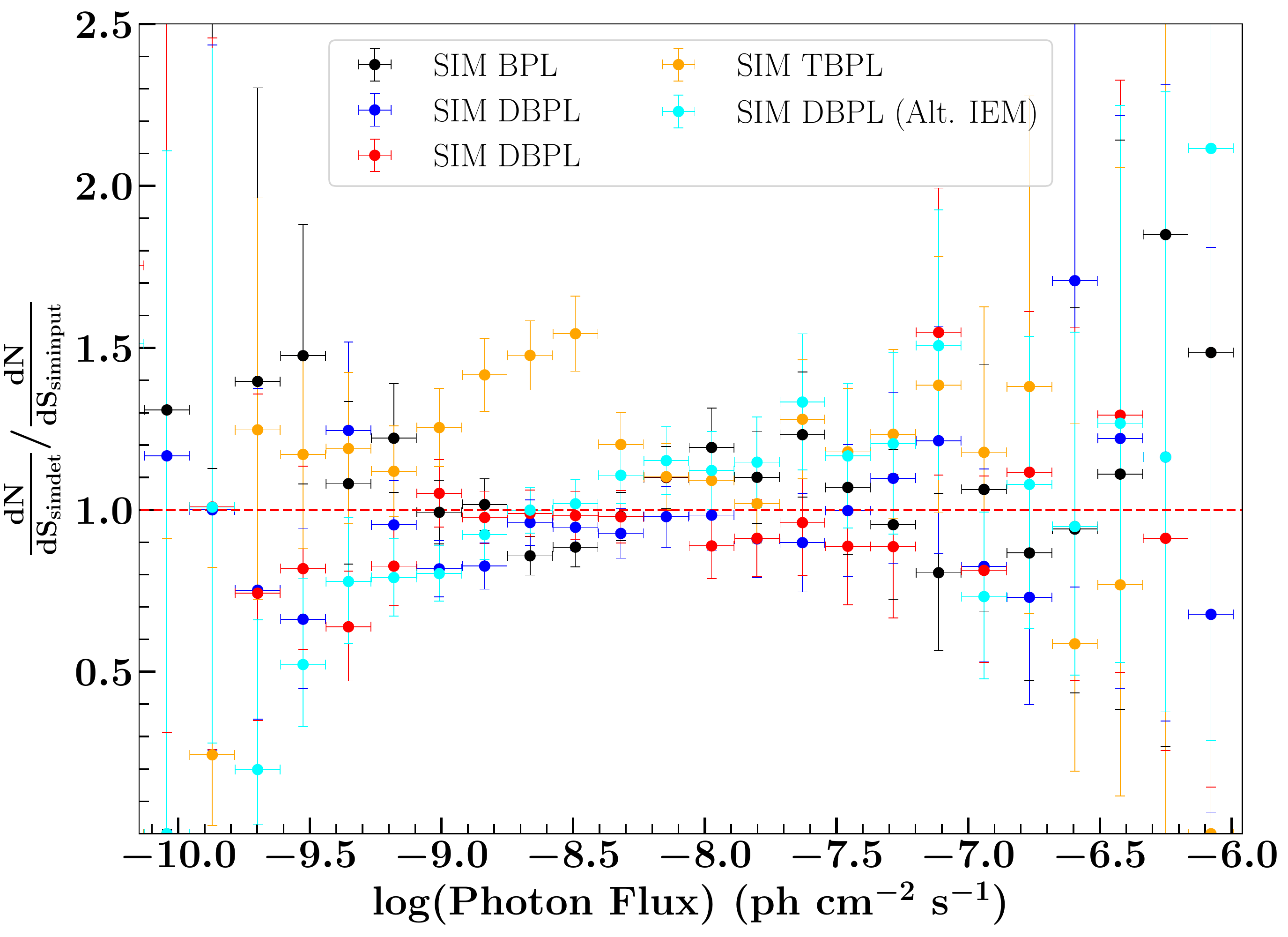}
\includegraphics[width=0.97\columnwidth]{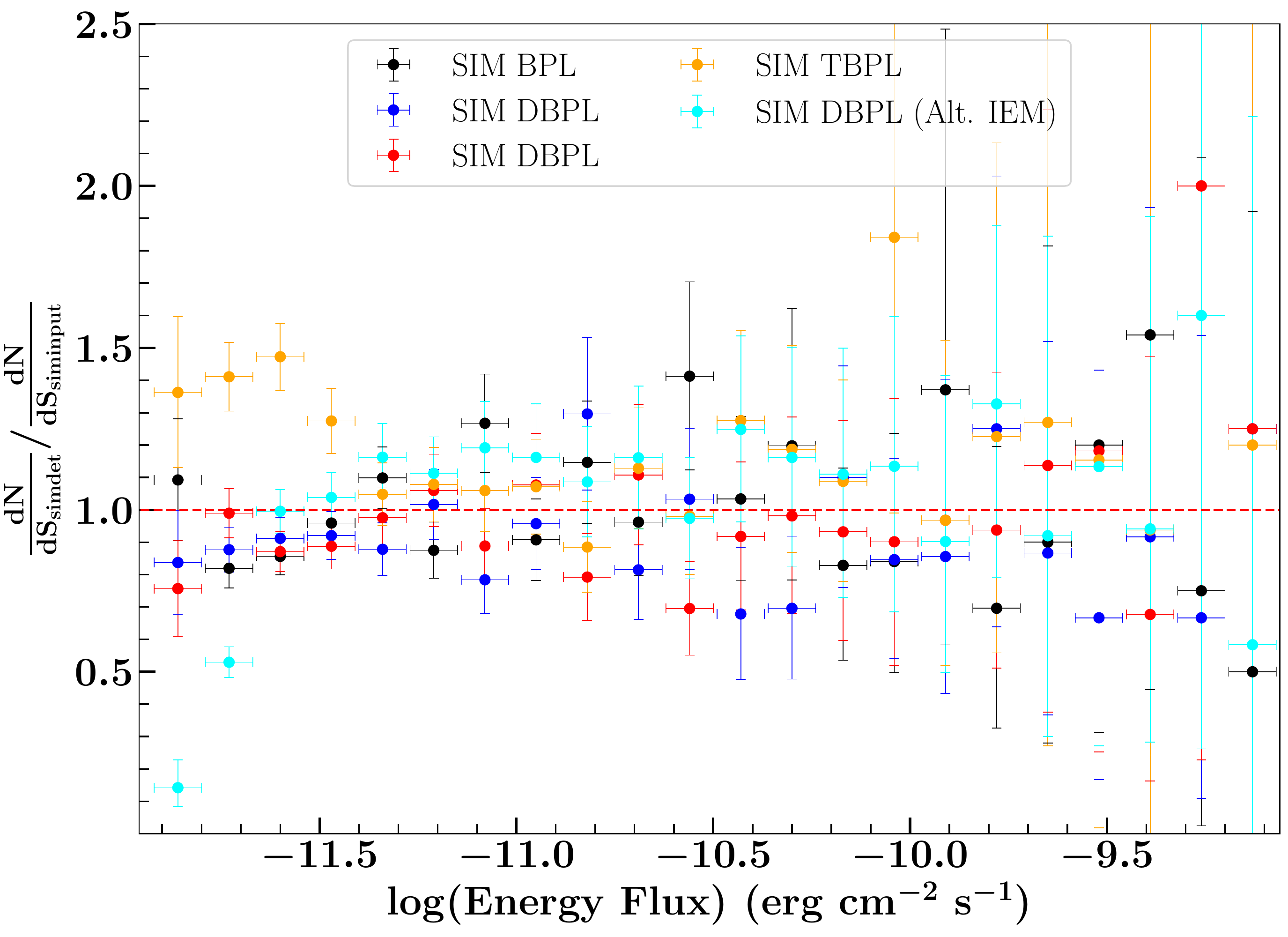}
	\caption{Ratios of the five simulated corrected {\it logN-logS} ($dN/dS_{\rm simdet}$) to the input ones ($dN/dS_{\rm siminput}$). The efficiency used to correct every simulation is the one derived from the combination of the other four, in order to  assess the systematics of the analysis. On the top we plot them for photon flux and on the bottom for energy flux. The systematics are taken, for every flux bin, as the lowest and highest ratio values.\label{fig:syst_r}}
\end{figure}
With the goal of assessing the systematic uncertainties affecting our derived source-count distribution, we consider the systematics arising from different {\it logN-logS} shapes used as input for the simulations and the effect of an alternative IEM model. For the first point, we employ all four simulations obtained with shapes described in Section~\ref{sec:sim}. In the case of the alternative IEM, we perform a fifth simulation (following the guidelines described in Section~\ref{sec:sim}) adding the IEM used for the study of the Galactic center \citep[][]{2017ApJ...840...43A}.

We derive the systematics of every simulation by correcting the simulated detected catalog with an efficiency obtained combining the other four. The ratio of the inferred (simulated) {\it logN-logS} ($dN/dS_{\rm simdet}$) to the input of the simulation ($dN/dS_{\rm siminput}$) gives us 
an understanding of the importance of the choice of the precise shape of the {\it logN-logS} input into the simulations.
 The results for both photon and energy flux are shown in Figure~\ref{fig:syst_r}. The total systematics are evaluated as the lowest and highest ratio values for every flux bin. 
Finally, we propagate these systematics to the {\it logN-logS}. These results are shown in Figure~\ref{fig:lognlogs}. The systematic uncertainties are of the order $\sim$10\,\% for the majority of flux bins and they increase up to $70\%$ at the extreme of the flux distribution where the number of sources is very small. 
\begin{figure}
\centering
\includegraphics[width=1.0\columnwidth]{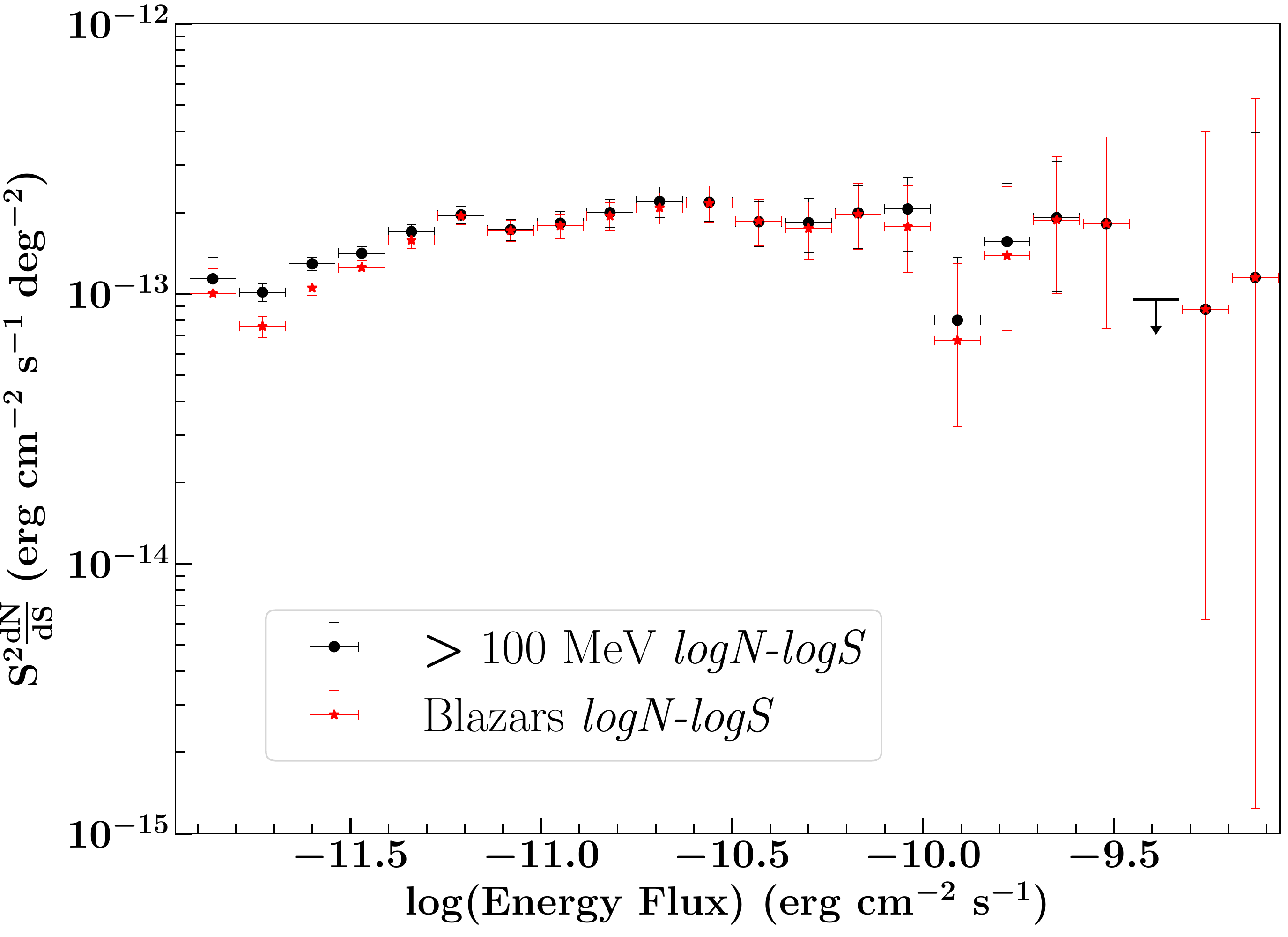}	
	\caption{Total {\it logN-logS} (black data points) and blazars {\it logN-logS} (red data points). Within errors, the two distributions are in very good agreement, showing that the {\it logN-logS} we derive in our analysis is indeed the blazars' one. \label{fig:complete}}
\end{figure}

\subsection{Blazars' logN-logS}\label{sec:complete}
We can derive the blazars' {\it logN-logS} using only the sources associated with blazars and correcting for the incompleteness of the associations, i.e. accounting for the number of blazars hiding among the unassociated sources.
Following the association listed in the FL8Y, our catalog contains 1906 blazars out of the 2680 detected sources. The incompleteness of the sample is defined as the sum of all sources matching with an unassociated FL8Y source, and sources without a positional counterpart. This results in 692 sources, which amounts to $\sim25\%$ of our sample.
Following a reasonable zeroth-order assumption, these objects are likely to be distributed in source classes similar to the associated portion of the sample. In the FL8Y, at $|b|>20\degree$, $\sim95.6\%$ of the associations are blazars, $\sim2.2\%$ pulsars and $\sim2.2\%$ belong to other classes (e.g.,\ AGN, globular clusters, etc.). Assuming these ratios\footnote{These ratios for blazars are always within 95\,\% and 100\,\% in every flux bin.} for the unassociated sources, we can derive the likely number of blazars among the unassociated sources as a function of flux.
Next, we can add these counts to the number per flux bin of classified blazars and derive the {\it total} blazar source-count distribution. We show this result in Figure~\ref{fig:complete}, where we compare it to the {\it logN-logS} of the extragalactic sky. It is evident how the two distributions are perfectly compatible, within errors.  

\section{Contribution to the EGB and Evolution of blazars}\label{sec:disc}
Having derived an improved {\it logN-logS} above $100\rm\,MeV$, as well as the detection efficiency, we can derive the contribution of blazars to the whole EGB ($S_{\rm EGB}$). We first use the method employed by \citet{2018ApJ...856..106D}:
\begin{equation}
	S_{\rm EGB}= \Sigma_{i=1}^N \frac{S_{PS, i}}{\Omega} + \int_{S_{\rm min}}^{S_{\rm max}} (1-\omega(S'))S'\frac{dN}{dS'}dS' \label{eq:contr}
\end{equation}
where $S_{PS, i}$ is the flux of the point source $i$ detected in our real catalog. The total flux of resolved point sources is summed to the flux of unresolved ones, obtained by integrating the best-fit {\it logN-logS} shape in the unresolved regime and taking into account the efficiency. The chosen integral limits are $10^{-11}\phflux$ and $10^{-6}\phflux$, the minimum and maximum source photon flux detected in our real catalog.
The result using the PL shape is $S_{\rm EGB}=5.60^{+0.95}_{-0.45}\times10^{-6}\phflux \rm sr^{-1}$, which implies that blazars contribute $50^{+10}_{-5}\%$ to the EGB. We note that the contribution of the unresolved point source (second term of Equation~\ref{eq:contr}) is $1.33^{+0.85}_{-0.14}\times10^{-6}\phflux\rm sr^{-1}$, which amounts to $\sim$27\% of the total contribution of blazars to the EGB.

The newly determined {\it logN-logS} also constrains the evolution of blazars  effectively. In \citet{2015ApJ...800L..27A}, three different models of blazar evolution were proposed and were found to explain the properties of blazars well. However, it was not possible to determine which model was the most representative of the blazar population. For example, it was not clear whether blazars were experiencing evolution primarily in luminosity (PLE model), or in density (PDE model), or in both density and luminosity (LDDE model). In this work we find that the LDDE and PLE models do not reproduce the faint end of the {\it logN-logS}, predicting less sources than observed. On the other hand, the PDE model reproduces the photon flux and energy flux {\it logN-logS} particularly well, now becoming the model of choice for the blazar evolution. This can be seen in see Figure~\ref{fig:lognlogs}, where the cyan, red and orange shaded areas highlight the different model predictions.
The PDE model allows us to get a second (complementary) estimate on the contribution of blazars to the EGB. This results in $S_{\rm EGB}=6.90^{+2.27}_{-0.6}\times10^{-6}\phflux\rm sr^{-1}$, which corresponds to $61^{+20}_{-6}\%$ of the total EGB, a value perfectly compatible with the range obtained by the first method (above). 

\section{Summary and Conclusions}
Blazars are the most abundant source class detected by the LAT. However, their contribution to the entire EGB is still an open issue. Previous studies have found that they can only contribute up to $50_{-11}^{+12}\%$ of the total EGB, and are not able to explain the IGRB below 100 GeV \citep[see e.g.,][]{2015PhRvD..91l3001D, 2015ApJ...799...86A, 2015ApJ...800L..27A,2018ApJ...856..106D}.  
In this work we derive the deepest source-count distribution to date, exploiting Fermi data above $\rm 100\,MeV$. Using 8 years of LAT data and the excellent quality of the Pass 8 dataset, we are able to resolve this distribution down to $10^{-10}$\phflux ($10^{-12}\ergflux$), an order of magnitude below the previous measurement by \citet{2010ApJ...720..435A}. In our analysis we employ the {\it efficiency correction} method, which allows us to account for the survey and data analysis biases. These results enable us to quantify the contribution of blazars to the total EGB, which is $\sim 50^{+10}_{-5}\%$. Furthermore, comparing our derived distribution with models of blazar evolution, we are also able to discern a favorable evolution model for this population. We find that, among the three models proposed in \citet{2015ApJ...800L..27A}, only the PDE model can reproduce our new {\it logN-logS} distribution. This implies a blazar contribution to the total background of $60^{+20}_{-6}\%$, compatible with the range obtained by our analysis. Our prediction is consistent with previous ones, confirming that the blazar population does not account for the total EGB. The remainder of this emission is likely attributed to other \gm-ray emitting source classes, such as MAGNs and starburst galaxies \citep[see e.g.,][]{2011ApJ...733...66I, 2014ApJ...780..161D,2015ApJ...800L..27A}. 

\acknowledgments
The \textit{Fermi} LAT Collaboration acknowledges generous ongoing support
from a number of agencies and institutes that have supported both the
development and the operation of the LAT as well as scientific data analysis.
These include the National Aeronautics and Space Administration and the
Department of Energy in the United States, the Commissariat \`a l'Energie Atomique
and the Centre National de la Recherche Scientifique / Institut National de Physique
Nucl\'eaire et de Physique des Particules in France, the Agenzia Spaziale Italiana
and the Istituto Nazionale di Fisica Nucleare in Italy, the Ministry of Education,
Culture, Sports, Science and Technology (MEXT), High Energy Accelerator Research
Organization (KEK) and Japan Aerospace Exploration Agency (JAXA) in Japan, and
the K.~A.~Wallenberg Foundation, the Swedish Research Council and the
Swedish National Space Board in Sweden.
 
Additional support for science analysis during the operations phase is gratefully
acknowledged from the Istituto Nazionale di Astrofisica in Italy and the Centre
National d'\'Etudes Spatiales in France. This work performed in part under DOE
Contract DE-AC02-76SF00515.
\bibliographystyle{aasjournal}

\bibliography{LogNLogS}

\begin{thebibliography}{}
\expandafter\ifx\csname natexlab\endcsname\relax\def\natexlab#1{#1}\fi

\bibitem[{{Abdo} {et~al.}(2010{\natexlab{a}}){Abdo}, {Ackermann}, {Ajello},
  {Allafort}, {Antolini}, {Atwood}, {Axelsson}, {Baldini}, {Ballet},
  {Barbiellini}, {Bastieri}, {Baughman}, {Bechtol}, {Bellazzini}, {Belli},
  {Berenji}, {Bisello}, {Blandford}, {Bloom}, {Bonamente}, {Bonnell},
  {Borgland}, {Bouvier}, {Bregeon}, {Brez}, {Brigida}, {Bruel}, {Burnett},
  {Busetto}, {Buson}, {Caliandro}, {Cameron}, {Campana}, {Canadas}, {Caraveo},
  {Carrigan}, {Casandjian}, {Cavazzuti}, {Ceccanti}, {Cecchi}, {{\c{C}}elik},
  {Charles}, {Chekhtman}, {Cheung}, {Chiang}, {Cillis}, {Ciprini}, {Claus},
  {Cohen-Tanugi}, {Conrad}, {Corbet}, {Davis}, {DeKlotz}, {den Hartog},
  {Dermer}, {de Angelis}, {de Luca}, {de Palma}, {Digel}, {Dormody}, {Silva},
  {Drell}, {Dubois}, {Dumora}, {Fabiani}, {Farnier}, {Favuzzi}, {Fegan},
  {Ferrara}, {Focke}, {Fortin}, {Frailis}, {Fukazawa}, {Funk}, {Fusco},
  {Gargano}, {Gasparrini}, {Gehrels}, {Germani}, {Giavitto}, {Giebels},
  {Giglietto}, {Giommi}, {Giordano}, {Giroletti}, {Glanzman}, {Godfrey},
  {Grenier}, {Grondin}, {Grove}, {Guillemot}, {Guiriec}, {Gustafsson},
  {Hadasch}, {Hanabata}, {Harding}, {Hayashida}, {Hays}, {Healey}, {Hill},
  {Horan}, {Hughes}, {Iafrate}, {J{\'o}hannesson}, {Johnson}, {Johnson},
  {Johnson}, {Johnson}, {Kamae}, {Katagiri}, {Kataoka}, {Kawai}, {Kerr},
  {Kn{\"o}dlseder}, {Kocevski}, {Kuss}, {Lande}, {Landriu}, {Latronico}, {Lee},
  {Lemoine-Goumard}, {Lionetto}, {Llena Garde}, {Longo}, {Loparco}, {Lott},
  {Lovellette}, {Lubrano}, {Madejski}, {Makeev}, {Marangelli}, {Marelli},
  {Massaro}, {Mazziotta}, {McConville}, {McEnery}, {Michelson}, {Minuti},
  {Mitthumsiri}, {Mizuno}, {Moiseev}, {Mongelli}, {Monte}, {Monzani},
  {Moretti}, {Morselli}, {Moskalenko}, {Murgia}, {Nakajima}, {Nakamori},
  {Naumann-Godo}, {Nolan}, {Norris}, {Nuss}, {Ohno}, {Ohsugi}, {Omodei},
  {Orlando}, {Ormes}, {Ozaki}, {Paccagnella}, {Paneque}, {Panetta}, {Parent},
  {Pelassa}, {Pepe}, {Pesce-Rollins}, {Pinchera}, {Piron}, {Porter}, {Poupard},
  {Rain{\`o}}, {Rando}, {Ray}, {Razzano}, {Razzaque}, {Rea}, {Reimer},
  {Reimer}, {Reposeur}, {Ripken}, {Ritz}, {Rochester}, {Rodriguez}, {Romani},
  {Roth}, {Sadrozinski}, {Salvetti}, {Sanchez}, {Sander}, {Saz Parkinson},
  {Scargle}, {Schalk}, {Scolieri}, {Sgr{\`o}}, {Shaw}, {Siskind}, {Smith},
  {Smith}, {Spandre}, {Spinelli}, {Starck}, {Stephens}, {Striani}, {Strickman},
  {Strong}, {Suson}, {Tajima}, {Takahashi}, {Takahashi}, {Tanaka}, {Thayer},
  {Thayer}, {Thompson}, {Tibaldo}, {Tibolla}, {Tinebra}, {Torres}, {Tosti},
  {Tramacere}, {Uchiyama}, {Usher}, {Van Etten}, {Vasileiou}, {Vilchez},
  {Vitale}, {Waite}, {Wallace}, {Wang}, {Watters}, {Winer}, {Wood}, {Yang},
  {Ylinen}, {Ziegler}, \& {Fermi LAT Collaboration}}]{2010ApJS..188..405A}
{Abdo}, A.~A., {Ackermann}, M., {Ajello}, M., {et~al.} 2010{\natexlab{a}}, The
  Astrophysical Journal Supplement Series, 188, 405

\bibitem[{{Abdo} {et~al.}(2010{\natexlab{b}}){Abdo}, {Ackermann}, {Ajello},
  {Baldini}, {Ballet}, {Barbiellini}, {Bastieri}, {Bechtol}, {Bellazzini},
  {Berenji}, {Blandford}, {Bloom}, {Bonamente}, {Borgland}, {Bouvier},
  {Brandt}, {Bregeon}, {Brez}, {Brigida}, {Bruel}, {Buehler}, {Burnett},
  {Buson}, {Caliandro}, {Cameron}, {Cannon}, {Caraveo}, {Carrigan},
  {Casandjian}, {Cavazzuti}, {Cecchi}, {{\c C}elik}, {Celotti}, {Charles},
  {Chekhtman}, {Chen}, {Cheung}, {Chiang}, {Ciprini}, {Claus}, {Cohen-Tanugi},
  {Colafrancesco}, {Conrad}, {Davis}, {Dermer}, {de Angelis}, {de Palma},
  {Silva}, {Drell}, {Dubois}, {Favuzzi}, {Fegan}, {Ferrara}, {Fortin},
  {Frailis}, {Fukazawa}, {Fusco}, {Gargano}, {Gasparrini}, {Gehrels},
  {Germani}, {Giglietto}, {Giommi}, {Giordano}, {Giroletti}, {Glanzman},
  {Godfrey}, {Grandi}, {Grenier}, {Grove}, {Guillemot}, {Guiriec}, {Hadasch},
  {Hayashida}, {Hays}, {Horan}, {Hughes}, {Jackson}, {J{\'o}hannesson},
  {Johnson}, {Johnson}, {Kamae}, {Katagiri}, {Kataoka}, {Kn{\"o}dlseder},
  {Kuss}, {Lande}, {Latronico}, {Lee}, {Lemoine-Goumard}, {Llena Garde},
  {Longo}, {Loparco}, {Lott}, {Lovellette}, {Lubrano}, {Madejski}, {Makeev},
  {Malaguti}, {Mazziotta}, {McConville}, {McEnery}, {Michelson}, {Migliori},
  {Mitthumsiri}, {Mizuno}, {Monte}, {Monzani}, {Morselli}, {Moskalenko},
  {Murgia}, {Naumann-Godo}, {Nestoras}, {Nolan}, {Norris}, {Nuss}, {Ohsugi},
  {Okumura}, {Omodei}, {Orlando}, {Ormes}, {Paneque}, {Panetta}, {Parent},
  {Pelassa}, {Pepe}, {Persic}, {Pesce-Rollins}, {Piron}, {Porter}, {Rain{\`o}},
  {Rando}, {Razzano}, {Razzaque}, {Reimer}, {Reimer}, {Reyes}, {Roth},
  {Sadrozinski}, {Sanchez}, {Sander}, {Scargle}, {Sgr{\`o}}, {Siskind},
  {Smith}, {Spandre}, {Spinelli}, {Stawarz}, {Stecker}, {Strickman}, {Suson},
  {Takahashi}, {Tanaka}, {Thayer}, {Thayer}, {Thompson}, {Tibaldo}, {Torres},
  {Torresi}, {Tosti}, {Tramacere}, {Uchiyama}, {Usher}, {Vandenbroucke},
  {Vasileiou}, {Vilchez}, {Villata}, {Vitale}, {Waite}, {Wang}, {Winer},
  {Wood}, {Yang}, {Ylinen}, \& {Ziegler}}]{2010ApJ...720..912A}
---. 2010{\natexlab{b}}, \apj, 720, 912

\bibitem[{{Abdo} {et~al.}(2010{\natexlab{c}}){Abdo}, {Ackermann}, {Ajello},
  {Antolini}, {Baldini}, {Ballet}, {Barbiellini}, {Bastieri}, {Baughman},
  {Bechtol}, {Bellazzini}, {Berenji}, {Blandford}, {Bloom}, {Bonamente},
  {Borgland}, {Bouvier}, {Bregeon}, {Brez}, {Brigida}, {Bruel}, {Burnett},
  {Buson}, {Caliandro}, {Cameron}, {Caraveo}, {Carrigan}, {Casandjian},
  {Cavazzuti}, {Cecchi}, {{\c C}elik}, {Charles}, {Chekhtman}, {Cheung},
  {Chiang}, {Ciprini}, {Claus}, {Cohen-Tanugi}, {Conrad}, {Costamante},
  {Cutini}, {Dermer}, {de Angelis}, {de Palma}, {Silva}, {Drell}, {Dubois},
  {Dumora}, {Farnier}, {Favuzzi}, {Fegan}, {Focke}, {Fukazawa}, {Funk},
  {Fusco}, {Gargano}, {Gasparrini}, {Gehrels}, {Germani}, {Giglietto},
  {Giommi}, {Giordano}, {Glanzman}, {Godfrey}, {Grenier}, {Grove}, {Guiriec},
  {Hadasch}, {Hayashida}, {Hays}, {Healey}, {Horan}, {Hughes}, {Itoh},
  {J{\'o}hannesson}, {Johnson}, {Johnson}, {Johnson}, {Kamae}, {Katagiri},
  {Kataoka}, {Kawai}, {Kn{\"o}dlseder}, {Kuss}, {Lande}, {Latronico}, {Lee},
  {Lemoine-Goumard}, {Llena Garde}, {Longo}, {Loparco}, {Lott}, {Lovellette},
  {Lubrano}, {Madejski}, {Makeev}, {Mazziotta}, {McConville}, {McEnery},
  {Meurer}, {Michelson}, {Mitthumsiri}, {Mizuno}, {Monte}, {Monzani},
  {Morselli}, {Moskalenko}, {Murgia}, {Nolan}, {Norris}, {Nuss}, {Ohsugi},
  {Omodei}, {Orlando}, {Ormes}, {Ozaki}, {Paneque}, {Panetta}, {Parent},
  {Pelassa}, {Pepe}, {Pesce-Rollins}, {Piron}, {Porter}, {Rain{\`o}}, {Rando},
  {Razzano}, {Reimer}, {Reimer}, {Ritz}, {Rochester}, {Rodriguez}, {Romani},
  {Roth}, {Sadrozinski}, {Sander}, {Saz Parkinson}, {Scargle}, {Sgr{\`o}},
  {Shaw}, {Smith}, {Spandre}, {Spinelli}, {Starck}, {Strickman}, {Strong},
  {Suson}, {Tajima}, {Takahashi}, {Takahashi}, {Tanaka}, {Thayer}, {Thayer},
  {Thompson}, {Tibaldo}, {Torres}, {Tosti}, {Tramacere}, {Uchiyama}, {Usher},
  {Vasileiou}, {Vilchez}, {Vitale}, {Waite}, {Wang}, {Winer}, {Wood}, {Yang},
  {Ylinen}, {Ziegler}, \& {Fermi LAT Collaboration}}]{2010ApJ...720..435A}
---. 2010{\natexlab{c}}, \apj, 720, 435

\bibitem[{{Abdollahi} {et~al.}(2020){Abdollahi}, {Acero}, {Ackermann},
  {Ajello}, {Atwood}, {Axelsson}, {Baldini}, {Ballet}, {Barbiellini},
  {Bastieri}, {Becerra Gonzalez}, {Bellazzini}, {Berretta}, {Bissaldi}, {Bland
  ford}, {Bloom}, {Bonino}, {Bottacini}, {Brandt}, {Bregeon}, {Bruel},
  {Buehler}, {Burnett}, {Buson}, {Cameron}, {Caputo}, {Caraveo}, {Casandjian},
  {Castro}, {Cavazzuti}, {Charles}, {Chaty}, {Chen}, {Cheung}, {Chiaro},
  {Ciprini}, {Cohen-Tanugi}, {Cominsky}, {Coronado-Bl{\'a}zquez}, {Costantin},
  {Cuoco}, {Cutini}, {D'Ammando}, {DeKlotz}, {Torre Luque}, {de Palma},
  {Desai}, {Digel}, {Lalla}, {Mauro}, {Venere}, {Dom{\'\i}nguez}, {Dumora},
  {Dirirsa}, {Fegan}, {Ferrara}, {Franckowiak}, {Fukazawa}, {Funk}, {Fusco},
  {Gargano}, {Gasparrini}, {Giglietto}, {Giommi}, {Giordano}, {Giroletti},
  {Glanzman}, {Green}, {Grenier}, {Griffin}, {Grondin}, {Grove}, {Guiriec},
  {Harding}, {Hayashi}, {Hays}, {Hewitt}, {Horan}, {J{\'o}hannesson},
  {Johnson}, {Kamae}, {Kerr}, {Kocevski}, {Kovac'evic'}, {Kuss}, {Landriu},
  {Larsson}, {Latronico}, {Lemoine-Goumard}, {Li}, {Liodakis}, {Longo},
  {Loparco}, {Lott}, {Lovellette}, {Lubrano}, {Madejski}, {Maldera},
  {Malyshev}, {Manfreda}, {Marchesini}, {Marcotulli}, {Mart{\'\i}-Devesa},
  {Martin}, {Massaro}, {Mazziotta}, {McEnery}, {Mereu}, {Meyer}, {Michelson},
  {Mirabal}, {Mizuno}, {Monzani}, {Morselli}, {Moskalenko}, {Negro}, {Nuss},
  {Ojha}, {Omodei}, {Orienti}, {Orlando}, {Ormes}, {Palatiello}, {Paliya},
  {Paneque}, {Pei}, {Pe{\~n}a-Herazo}, {Perkins}, {Persic}, {Pesce-Rollins},
  {Petrosian}, {Petrov}, {Piron}, {Poon}, {Porter}, {Principe}, {Rain{\`o}},
  {Rando}, {Razzano}, {Razzaque}, {Reimer}, {Reimer}, {Remy}, {Reposeur},
  {Romani}, {Parkinson}, {Schinzel}, {Serini}, {Sgr{\`o}}, {Siskind}, {Smith},
  {Spandre}, {Spinelli}, {Strong}, {Suson}, {Tajima}, {Takahashi}, {Tak},
  {Thayer}, {Thompson}, {Tibaldo}, {Torres}, {Torresi}, {Valverde}, {Klaveren},
  {Zyl}, {Wood}, {Yassine}, \& {Zaharijas}}]{2020ApJS..247...33A}
{Abdollahi}, S., {Acero}, F., {Ackermann}, M., {et~al.} 2020, \apjs, 247, 33

\bibitem[{{Acero} {et~al.}(2015){Acero}, {Ackermann}, {Ajello}, {Albert},
  {Atwood}, {Axelsson}, {Baldini}, {Ballet}, {Barbiellini}, {Bastieri},
  {Belfiore}, {Bellazzini}, {Bissaldi}, {Blandford}, {Bloom}, {Bogart},
  {Bonino}, {Bottacini}, {Bregeon}, {Britto}, {Bruel}, {Buehler}, {Burnett},
  {Buson}, {Caliandro}, {Cameron}, {Caputo}, {Caragiulo}, {Caraveo},
  {Casandjian}, {Cavazzuti}, {Charles}, {Chaves}, {Chekhtman}, {Cheung},
  {Chiang}, {Chiaro}, {Ciprini}, {Claus}, {Cohen-Tanugi}, {Cominsky}, {Conrad},
  {Cutini}, {D'Ammando}, {de Angelis}, {DeKlotz}, {de Palma}, {Desiante},
  {Digel}, {Di Venere}, {Drell}, {Dubois}, {Dumora}, {Favuzzi}, {Fegan},
  {Ferrara}, {Finke}, {Franckowiak}, {Fukazawa}, {Funk}, {Fusco}, {Gargano},
  {Gasparrini}, {Giebels}, {Giglietto}, {Giommi}, {Giordano}, {Giroletti},
  {Glanzman}, {Godfrey}, {Grenier}, {Grondin}, {Grove}, {Guillemot}, {Guiriec},
  {Hadasch}, {Harding}, {Hays}, {Hewitt}, {Hill}, {Horan}, {Iafrate}, {Jogler},
  {J{\'o}hannesson}, {Johnson}, {Johnson}, {Johnson}, {Johnson}, {Kamae},
  {Kataoka}, {Katsuta}, {Kuss}, {La Mura}, {Landriu}, {Larsson}, {Latronico},
  {Lemoine-Goumard}, {Li}, {Li}, {Longo}, {Loparco}, {Lott}, {Lovellette},
  {Lubrano}, {Madejski}, {Massaro}, {Mayer}, {Mazziotta}, {McEnery},
  {Michelson}, {Mirabal}, {Mizuno}, {Moiseev}, {Mongelli}, {Monzani},
  {Morselli}, {Moskalenko}, {Murgia}, {Nuss}, {Ohno}, {Ohsugi}, {Omodei},
  {Orienti}, {Orlando}, {Ormes}, {Paneque}, {Panetta}, {Perkins},
  {Pesce-Rollins}, {Piron}, {Pivato}, {Porter}, {Racusin}, {Rando}, {Razzano},
  {Razzaque}, {Reimer}, {Reimer}, {Reposeur}, {Rochester}, {Romani},
  {Salvetti}, {S{\'a}nchez-Conde}, {Saz Parkinson}, {Schulz}, {Siskind},
  {Smith}, {Spada}, {Spandre}, {Spinelli}, {Stephens}, {Strong}, {Suson},
  {Takahashi}, {Takahashi}, {Tanaka}, {Thayer}, {Thayer}, {Thompson},
  {Tibaldo}, {Tibolla}, {Torres}, {Torresi}, {Tosti}, {Troja}, {Van Klaveren},
  {Vianello}, {Winer}, {Wood}, {Wood}, {Zimmer}, \& {Fermi-LAT
  Collaboration}}]{2015ApJS..218...23A}
{Acero}, F., {Ackermann}, M., {Ajello}, M., {et~al.} 2015, The Astrophysical
  Journal Supplement Series, 218, 23

\bibitem[{{Acero} {et~al.}(2016){Acero}, {Ackermann}, {Ajello}, {Albert},
  {Baldini}, {Ballet}, {Barbiellini}, {Bastieri}, {Bellazzini}, {Bissaldi},
  {Bloom}, {Bonino}, {Bottacini}, {Brandt}, {Bregeon}, {Bruel}, {Buehler},
  {Buson}, {Caliandro}, {Cameron}, {Caragiulo}, {Caraveo}, {Casandjian},
  {Cavazzuti}, {Cecchi}, {Charles}, {Chekhtman}, {Chiang}, {Chiaro}, {Ciprini},
  {Claus}, {Cohen-Tanugi}, {Conrad}, {Cuoco}, {Cutini}, {D'Ammando}, {de
  Angelis}, {de Palma}, {Desiante}, {Digel}, {Di Venere}, {Drell}, {Favuzzi},
  {Fegan}, {Ferrara}, {Focke}, {Franckowiak}, {Funk}, {Fusco}, {Gargano},
  {Gasparrini}, {Giglietto}, {Giordano}, {Giroletti}, {Glanzman}, {Godfrey},
  {Grenier}, {Guiriec}, {Hadasch}, {Harding}, {Hayashi}, {Hays}, {Hewitt},
  {Hill}, {Horan}, {Hou}, {Jogler}, {J{\'o}hannesson}, {Kamae}, {Kuss},
  {Landriu}, {Larsson}, {Latronico}, {Li}, {Li}, {Longo}, {Loparco},
  {Lovellette}, {Lubrano}, {Maldera}, {Malyshev}, {Manfreda}, {Martin},
  {Mayer}, {Mazziotta}, {McEnery}, {Michelson}, {Mirabal}, {Mizuno}, {Monzani},
  {Morselli}, {Nuss}, {Ohsugi}, {Omodei}, {Orienti}, {Orlando}, {Ormes},
  {Paneque}, {Pesce-Rollins}, {Piron}, {Pivato}, {Rain{\`o}}, {Rando},
  {Razzano}, {Razzaque}, {Reimer}, {Reimer}, {Remy}, {Renault},
  {S{\'a}nchez-Conde}, {Schaal}, {Schulz}, {Sgr{\`o}}, {Siskind}, {Spada},
  {Spandre}, {Spinelli}, {Strong}, {Suson}, {Tajima}, {Takahashi}, {Thayer},
  {Thompson}, {Tibaldo}, {Tinivella}, {Torres}, {Tosti}, {Troja}, {Vianello},
  {Werner}, {Wood}, {Wood}, {Zaharijas}, \& {Zimmer}}]{2016ApJS..223...26A}
---. 2016, \apjs, 223, 26

\bibitem[{{Ackermann} {et~al.}(2012){Ackermann}, {Ajello}, {Allafort},
  {Baldini}, {Ballet}, {Bastieri}, {Bechtol}, {Bellazzini}, {Berenji}, {Bloom},
  {Bonamente}, {Borgland}, {Bouvier}, {Bregeon}, {Brigida}, {Bruel}, {Buehler},
  {Buson}, {Caliandro}, {Cameron}, {Caraveo}, {Casandjian}, {Cecchi},
  {Charles}, {Chekhtman}, {Cheung}, {Chiang}, {Cillis}, {Ciprini}, {Claus},
  {Cohen-Tanugi}, {Conrad}, {Cutini}, {de Palma}, {Dermer}, {Digel}, {Silva},
  {Drell}, {Drlica-Wagner}, {Favuzzi}, {Fegan}, {Fortin}, {Fukazawa}, {Funk},
  {Fusco}, {Gargano}, {Gasparrini}, {Germani}, {Giglietto}, {Giordano},
  {Glanzman}, {Godfrey}, {Grenier}, {Guiriec}, {Gustafsson}, {Hadasch},
  {Hayashida}, {Hays}, {Hughes}, {J{\'o}hannesson}, {Johnson}, {Kamae},
  {Katagiri}, {Kataoka}, {Kn{\"o}dlseder}, {Kuss}, {Lande}, {Longo}, {Loparco},
  {Lott}, {Lovellette}, {Lubrano}, {Madejski}, {Martin}, {Mazziotta},
  {McEnery}, {Michelson}, {Mizuno}, {Monte}, {Monzani}, {Morselli},
  {Moskalenko}, {Murgia}, {Nishino}, {Norris}, {Nuss}, {Ohno}, {Ohsugi},
  {Okumura}, {Omodei}, {Orlando}, {Ozaki}, {Parent}, {Persic}, {Pesce-Rollins},
  {Petrosian}, {Pierbattista}, {Piron}, {Pivato}, {Porter}, {Rain{\`o}},
  {Rando}, {Razzano}, {Reimer}, {Reimer}, {Ritz}, {Roth}, {Sbarra}, {Sgr{\`o}},
  {Siskind}, {Spandre}, {Spinelli}, {Stawarz}, {Strong}, {Takahashi}, {Tanaka},
  {Thayer}, {Tibaldo}, {Tinivella}, {Torres}, {Tosti}, {Troja}, {Uchiyama},
  {Vandenbroucke}, {Vianello}, {Vitale}, {Waite}, {Wood}, \&
  {Yang}}]{2012ApJ...755..164A}
{Ackermann}, M., {Ajello}, M., {Allafort}, A., {et~al.} 2012, \apj, 755, 164

\bibitem[{{Ackermann} {et~al.}(2015{\natexlab{a}}){Ackermann}, {Ajello},
  {Albert}, {Atwood}, {Baldini}, {Ballet}, {Barbiellini}, {Bastieri},
  {Bechtol}, {Bellazzini}, {Bissaldi}, {Blandford}, {Bloom}, {Bottacini},
  {Brandt}, {Bregeon}, {Bruel}, {Buehler}, {Buson}, {Caliandro}, {Cameron},
  {Caragiulo}, {Caraveo}, {Cavazzuti}, {Cecchi}, {Charles}, {Chekhtman},
  {Chiang}, {Chiaro}, {Ciprini}, {Claus}, {Cohen-Tanugi}, {Conrad}, {Cuoco},
  {Cutini}, {D'Ammando}, {de Angelis}, {de Palma}, {Dermer}, {Digel}, {Silva},
  {Drell}, {Favuzzi}, {Ferrara}, {Focke}, {Franckowiak}, {Fukazawa}, {Funk},
  {Fusco}, {Gargano}, {Gasparrini}, {Germani}, {Giglietto}, {Giommi},
  {Giordano}, {Giroletti}, {Godfrey}, {Gomez-Vargas}, {Grenier}, {Guiriec},
  {Gustafsson}, {Hadasch}, {Hayashi}, {Hays}, {Hewitt}, {Ippoliti}, {Jogler},
  {J{\'o}hannesson}, {Johnson}, {Johnson}, {Kamae}, {Kataoka},
  {Kn{\"o}dlseder}, {Kuss}, {Larsson}, {Latronico}, {Li}, {Li}, {Longo},
  {Loparco}, {Lott}, {Lovellette}, {Lubrano}, {Madejski}, {Manfreda},
  {Massaro}, {Mayer}, {Mazziotta}, {McEnery}, {Michelson}, {Mitthumsiri},
  {Mizuno}, {Moiseev}, {Monzani}, {Morselli}, {Moskalenko}, {Murgia}, {Nemmen},
  {Nuss}, {Ohsugi}, {Omodei}, {Orlando}, {Ormes}, {Paneque}, {Panetta},
  {Perkins}, {Pesce-Rollins}, {Piron}, {Pivato}, {Porter}, {Rain{\`o}},
  {Rando}, {Razzano}, {Razzaque}, {Reimer}, {Reimer}, {Reposeur}, {Ritz},
  {Romani}, {S{\'a}nchez-Conde}, {Schaal}, {Schulz}, {Sgr{\`o}}, {Siskind},
  {Spandre}, {Spinelli}, {Strong}, {Suson}, {Takahashi}, {Thayer}, {Thayer},
  {Tibaldo}, {Tinivella}, {Torres}, {Tosti}, {Troja}, {Uchiyama}, {Vianello},
  {Werner}, {Winer}, {Wood}, {Wood}, {Zaharijas}, \&
  {Zimmer}}]{2015ApJ...799...86A}
{Ackermann}, M., {Ajello}, M., {Albert}, A., {et~al.} 2015{\natexlab{a}}, \apj,
  799, 86

\bibitem[{{Ackermann} {et~al.}(2015{\natexlab{b}}){Ackermann}, {Ajello},
  {Atwood}, {Baldini}, {Ballet}, {Barbiellini}, {Bastieri}, {Becerra Gonzalez},
  {Bellazzini}, {Bissaldi}, {Blandford}, {Bloom}, {Bonino}, {Bottacini},
  {Brandt}, {Bregeon}, {Britto}, {Bruel}, {Buehler}, {Buson}, {Caliandro},
  {Cameron}, {Caragiulo}, {Caraveo}, {Carpenter}, {Casandjian}, {Cavazzuti},
  {Cecchi}, {Charles}, {Chekhtman}, {Cheung}, {Chiang}, {Chiaro}, {Ciprini},
  {Claus}, {Cohen-Tanugi}, {Cominsky}, {Conrad}, {Cutini}, {D'Abrusco},
  {D'Ammando}, {de Angelis}, {Desiante}, {Digel}, {Di Venere}, {Drell},
  {Favuzzi}, {Fegan}, {Ferrara}, {Finke}, {Focke}, {Franckowiak}, {Fuhrmann},
  {Fukazawa}, {Furniss}, {Fusco}, {Gargano}, {Gasparrini}, {Giglietto},
  {Giommi}, {Giordano}, {Giroletti}, {Glanzman}, {Godfrey}, {Grenier}, {Grove},
  {Guiriec}, {Hewitt}, {Hill}, {Horan}, {Itoh}, {J{\'o}hannesson}, {Johnson},
  {Johnson}, {Kataoka}, {Kawano}, {Krauss}, {Kuss}, {La Mura}, {Larsson},
  {Latronico}, {Leto}, {Li}, {Li}, {Longo}, {Loparco}, {Lott}, {Lovellette},
  {Lubrano}, {Madejski}, {Mayer}, {Mazziotta}, {McEnery}, {Michelson},
  {Mizuno}, {Moiseev}, {Monzani}, {Morselli}, {Moskalenko}, {Murgia}, {Nuss},
  {Ohno}, {Ohsugi}, {Ojha}, {Omodei}, {Orienti}, {Orlando}, {Paggi}, {Paneque},
  {Perkins}, {Pesce-Rollins}, {Piron}, {Pivato}, {Porter}, {Rain{\`o}},
  {Rando}, {Razzano}, {Razzaque}, {Reimer}, {Reimer}, {Romani}, {Salvetti},
  {Schaal}, {Schinzel}, {Schulz}, {Sgr{\`o}}, {Siskind}, {Sokolovsky}, {Spada},
  {Spandre}, {Spinelli}, {Stawarz}, {Suson}, {Takahashi}, {Takahashi},
  {Tanaka}, {Thayer}, {Thayer}, {Tibaldo}, {Torres}, {Torresi}, {Tosti},
  {Troja}, {Uchiyama}, {Vianello}, {Winer}, {Wood}, \&
  {Zimmer}}]{2015ApJ...810...14A}
{Ackermann}, M., {Ajello}, M., {Atwood}, W.~B., {et~al.} 2015{\natexlab{b}},
  \apj, 810, 14

\bibitem[{{Ackermann} {et~al.}(2016){Ackermann}, {Ajello}, {Albert}, {Atwood},
  {Baldini}, {Ballet}, {Barbiellini}, {Bastieri}, {Bechtol}, {Bellazzini},
  {Bissaldi}, {Blandford}, {Bloom}, {Bonino}, {Bregeon}, {Britto}, {Bruel},
  {Buehler}, {Caliandro}, {Cameron}, {Caragiulo}, {Caraveo}, {Cavazzuti},
  {Cecchi}, {Charles}, {Chekhtman}, {Chiang}, {Chiaro}, {Ciprini},
  {Cohen-Tanugi}, {Cominsky}, {Costanza}, {Cutini}, {D'Ammando}, {de Angelis},
  {de Palma}, {Desiante}, {Digel}, {Di Mauro}, {Di Venere}, {Dom{\'{\i}}nguez},
  {Drell}, {Favuzzi}, {Fegan}, {Ferrara}, {Franckowiak}, {Fukazawa}, {Funk},
  {Fusco}, {Gargano}, {Gasparrini}, {Giglietto}, {Giommi}, {Giordano},
  {Giroletti}, {Godfrey}, {Green}, {Grenier}, {Guiriec}, {Hays}, {Horan},
  {Iafrate}, {Jogler}, {J{\'o}hannesson}, {Kuss}, {La Mura}, {Larsson},
  {Latronico}, {Li}, {Li}, {Longo}, {Loparco}, {Lott}, {Lovellette}, {Lubrano},
  {Madejski}, {Magill}, {Maldera}, {Manfreda}, {Mayer}, {Mazziotta},
  {Michelson}, {Mitthumsiri}, {Mizuno}, {Moiseev}, {Monzani}, {Morselli},
  {Moskalenko}, {Murgia}, {Negro}, {Nuss}, {Ohsugi}, {Okada}, {Omodei},
  {Orlando}, {Ormes}, {Paneque}, {Perkins}, {Pesce-Rollins}, {Petrosian},
  {Piron}, {Pivato}, {Porter}, {Rain{\`o}}, {Rando}, {Razzano}, {Razzaque},
  {Reimer}, {Reimer}, {Reposeur}, {Romani}, {S{\'a}nchez-Conde}, {Schmid},
  {Schulz}, {Sgr{\`o}}, {Simone}, {Siskind}, {Spada}, {Spandre}, {Spinelli},
  {Suson}, {Takahashi}, {Thayer}, {Tibaldo}, {Torres}, {Troja}, {Vianello},
  {Yassine}, \& {Zimmer}}]{2016PhRvL.116o1105A}
{Ackermann}, M., {Ajello}, M., {Albert}, A., {et~al.} 2016, Physical Review
  Letters, 116, 151105

\bibitem[{{Ackermann} {et~al.}(2017){Ackermann}, {Ajello}, {Albert}, {Atwood},
  {Baldini}, {Ballet}, {Barbiellini}, {Bastieri}, {Bellazzini}, {Bissaldi},
  {Blandford}, {Bloom}, {Bonino}, {Bottacini}, {Brandt}, {Bregeon}, {Bruel},
  {Buehler}, {Burnett}, {Cameron}, {Caputo}, {Caragiulo}, {Caraveo},
  {Cavazzuti}, {Cecchi}, {Charles}, {Chekhtman}, {Chiang}, {Chiappo}, {Chiaro},
  {Ciprini}, {Conrad}, {Costanza}, {Cuoco}, {Cutini}, {D'Ammando}, {de Palma},
  {Desiante}, {Digel}, {Di Lalla}, {Di Mauro}, {Di Venere}, {Drell}, {Favuzzi},
  {Fegan}, {Ferrara}, {Focke}, {Franckowiak}, {Fukazawa}, {Funk}, {Fusco},
  {Gargano}, {Gasparrini}, {Giglietto}, {Giordano}, {Giroletti}, {Glanzman},
  {Gomez-Vargas}, {Green}, {Grenier}, {Grove}, {Guillemot}, {Guiriec},
  {Gustafsson}, {Harding}, {Hays}, {Hewitt}, {Horan}, {Jogler}, {Johnson},
  {Kamae}, {Kocevski}, {Kuss}, {La Mura}, {Larsson}, {Latronico}, {Li},
  {Longo}, {Loparco}, {Lovellette}, {Lubrano}, {Magill}, {Maldera}, {Malyshev},
  {Manfreda}, {Martin}, {Mazziotta}, {Michelson}, {Mirabal}, {Mitthumsiri},
  {Mizuno}, {Moiseev}, {Monzani}, {Morselli}, {Negro}, {Nuss}, {Ohsugi},
  {Orienti}, {Orlando}, {Ormes}, {Paneque}, {Perkins}, {Persic},
  {Pesce-Rollins}, {Piron}, {Principe}, {Rain{\`o}}, {Rando}, {Razzano},
  {Razzaque}, {Reimer}, {Reimer}, {S{\'a}nchez-Conde}, {Sgr{\`o}}, {Simone},
  {Siskind}, {Spada}, {Spandre}, {Spinelli}, {Suson}, {Tajima}, {Tanaka},
  {Thayer}, {Tibaldo}, {Torres}, {Troja}, {Uchiyama}, {Vianello}, {Wood},
  {Wood}, {Zaharijas}, {Zimmer}, \& {Fermi LAT
  Collaboration}}]{2017ApJ...840...43A}
---. 2017, \apj, 840, 43

\bibitem[{{Ahn} {et~al.}(2007){Ahn}, {Bertone}, {Merritt}, \&
  {Zhang}}]{2007PhRvD..76b3517A}
{Ahn}, E.-J., {Bertone}, G., {Merritt}, D., \& {Zhang}, P. 2007, \prd, 76,
  023517

\bibitem[{{Ajello} {et~al.}(2015){Ajello}, {Gasparrini}, {S{\'a}nchez-Conde},
  {Zaharijas}, {Gustafsson}, {Cohen-Tanugi}, {Dermer}, {Inoue}, {Hartmann},
  {Ackermann}, {Bechtol}, {Franckowiak}, {Reimer}, {Romani}, \&
  {Strong}}]{2015ApJ...800L..27A}
{Ajello}, M., {Gasparrini}, D., {S{\'a}nchez-Conde}, M., {et~al.} 2015, \apjl,
  800, L27

\bibitem[{{Ajello} {et~al.}(2017){Ajello}, {Atwood}, {Baldini}, {Ballet},
  {Barbiellini}, {Bastieri}, {Bellazzini}, {Bissaldi}, {Blandford}, {Bloom},
  {Bonino}, {Bregeon}, {Britto}, {Bruel}, {Buehler}, {Buson}, {Cameron},
  {Caputo}, {Caragiulo}, {Caraveo}, {Cavazzuti}, {Cecchi}, {Charles},
  {Chekhtman}, {Cheung}, {Chiaro}, {Ciprini}, {Cohen}, {Costantin}, {Costanza},
  {Cuoco}, {Cutini}, {D'Ammando}, {de Palma}, {Desiante}, {Digel}, {Di Lalla},
  {Di Mauro}, {Di Venere}, {Dom{\'{\i}}nguez}, {Drell}, {Dumora}, {Favuzzi},
  {Fegan}, {Ferrara}, {Fortin}, {Franckowiak}, {Fukazawa}, {Funk}, {Fusco},
  {Gargano}, {Gasparrini}, {Giglietto}, {Giommi}, {Giordano}, {Giroletti},
  {Glanzman}, {Green}, {Grenier}, {Grondin}, {Grove}, {Guillemot}, {Guiriec},
  {Harding}, {Hays}, {Hewitt}, {Horan}, {J{\'o}hannesson}, {Kensei}, {Kuss},
  {La Mura}, {Larsson}, {Latronico}, {Lemoine-Goumard}, {Li}, {Longo},
  {Loparco}, {Lott}, {Lubrano}, {Magill}, {Maldera}, {Manfreda}, {Mazziotta},
  {McEnery}, {Meyer}, {Michelson}, {Mirabal}, {Mitthumsiri}, {Mizuno},
  {Moiseev}, {Monzani}, {Morselli}, {Moskalenko}, {Negro}, {Nuss}, {Ohsugi},
  {Omodei}, {Orienti}, {Orlando}, {Palatiello}, {Paliya}, {Paneque}, {Perkins},
  {Persic}, {Pesce-Rollins}, {Piron}, {Porter}, {Principe}, {Rain{\`o}},
  {Rando}, {Razzano}, {Razzaque}, {Reimer}, {Reimer}, {Reposeur}, {Saz
  Parkinson}, {Sgr{\`o}}, {Simone}, {Siskind}, {Spada}, {Spandre}, {Spinelli},
  {Stawarz}, {Suson}, {Takahashi}, {Tak}, {Thayer}, {Thayer}, {Thompson},
  {Torres}, {Torresi}, {Troja}, {Vianello}, {Wood}, \&
  {Wood}}]{2017ApJS..232...18A}
{Ajello}, M., {Atwood}, W.~B., {Baldini}, L., {et~al.} 2017, \apjs, 232, 18

\bibitem[{{Ammazzalorso} {et~al.}(2018){Ammazzalorso}, {Fornengo}, {Horiuchi},
  \& {Regis}}]{2018PhRvD..98j3007A}
{Ammazzalorso}, S., {Fornengo}, N., {Horiuchi}, S., \& {Regis}, M. 2018, \prd,
  98, 103007

\bibitem[{{Ando} \& {Pavlidou}(2009)}]{2009MNRAS.400.2122A}
{Ando}, S., \& {Pavlidou}, V. 2009, \mnras, 400, 2122

\bibitem[{{Atwood} {et~al.}(2013){Atwood}, {Albert}, {Baldini}, {Tinivella},
  {Bregeon}, {Pesce-Rollins}, {Sgr{\`o}}, {Bruel}, {Charles}, {Drlica-Wagner},
  {Franckowiak}, {Jogler}, {Rochester}, {Usher}, {Wood}, {Cohen-Tanugi}, \&
  {S.~Zimmer for the Fermi-LAT Collaboration}}]{2013arXiv1303.3514A}
{Atwood}, W., {Albert}, A., {Baldini}, L., {et~al.} 2013, ArXiv e-prints,
  arXiv:1303.3514

\bibitem[{{Atwood} {et~al.}(2009){Atwood}, {Abdo}, {Ackermann}, {Althouse},
  {Anderson}, {Axelsson}, {Baldini}, {Ballet}, {Band}, {Barbiellini}, \&
  et~al.}]{2009ApJ...697.1071A}
{Atwood}, W.~B., {Abdo}, A.~A., {Ackermann}, M., {et~al.} 2009, \apj, 697, 1071

\bibitem[{{Bergstr{\"o}m} {et~al.}(2001){Bergstr{\"o}m}, {Edsj{\"o}}, \&
  {Ullio}}]{2001PhRvL..87y1301B}
{Bergstr{\"o}m}, L., {Edsj{\"o}}, J., \& {Ullio}, P. 2001, Physical Review
  Letters, 87, 251301

\bibitem[{{Bruel} {et~al.}(2018){Bruel}, {Burnett}, {Digel}, {Johannesson},
  {Omodei}, \& {Wood}}]{2018arXiv181011394B}
{Bruel}, P., {Burnett}, T.~H., {Digel}, S.~W., {et~al.} 2018, ArXiv e-prints,
  arXiv:1810.11394

\bibitem[{{Chakraborty} \& {Fields}(2013)}]{2013ApJ...773..104C}
{Chakraborty}, N., \& {Fields}, B.~D. 2013, \apj, 773, 104

\bibitem[{{Cuoco} {et~al.}(2017){Cuoco}, {Bilicki}, {Xia}, \&
  {Branchini}}]{2017ApJS..232...10C}
{Cuoco}, A., {Bilicki}, M., {Xia}, J.-Q., \& {Branchini}, E. 2017, The
  Astrophysical Journal Supplement Series, 232, 10

\bibitem[{{Di Mauro} {et~al.}(2014){Di Mauro}, {Calore}, {Donato}, {Ajello}, \&
  {Latronico}}]{2014ApJ...780..161D}
{Di Mauro}, M., {Calore}, F., {Donato}, F., {Ajello}, M., \& {Latronico}, L.
  2014, \apj, 780, 161

\bibitem[{{Di Mauro} \& {Donato}(2015)}]{2015PhRvD..91l3001D}
{Di Mauro}, M., \& {Donato}, F. 2015, \prd, 91, 123001

\bibitem[{{Di Mauro} {et~al.}(2018){Di Mauro}, {Manconi}, {Zechlin}, {Ajello},
  {Charles}, \& {Donato}}]{2018ApJ...856..106D}
{Di Mauro}, M., {Manconi}, S., {Zechlin}, H.-S., {et~al.} 2018, \apj, 856, 106

\bibitem[{{Eddington}(1913)}]{1913MNRAS..73..359E}
{Eddington}, A.~S. 1913, \mnras, 73, 359

\bibitem[{{Efron} \& {Petrosian}(1992)}]{1992ApJ...399..345E}
{Efron}, B., \& {Petrosian}, V. 1992, \apj, 399, 345

\bibitem[{{Fichtel} {et~al.}(1975){Fichtel}, {Hartman}, {Kniffen}, {Thompson},
  {Ogelman}, {Ozel}, {Tumer}, \& {Bignami}}]{1975ApJ...198..163F}
{Fichtel}, C.~E., {Hartman}, R.~C., {Kniffen}, D.~A., {et~al.} 1975, \apj, 198,
  163

\bibitem[{{Fields} {et~al.}(2010){Fields}, {Pavlidou}, \&
  {Prodanovi{\'c}}}]{2010ApJ...722L.199F}
{Fields}, B.~D., {Pavlidou}, V., \& {Prodanovi{\'c}}, T. 2010, \apjl, 722, L199

\bibitem[{{Fornasa} \& {S{\'a}nchez-Conde}(2015)}]{2015PhR...598....1F}
{Fornasa}, M., \& {S{\'a}nchez-Conde}, M.~A. 2015, \physrep, 598, 1

\bibitem[{{Inoue}(2011)}]{2011ApJ...733...66I}
{Inoue}, Y. 2011, \apj, 733, 66

\bibitem[{Kass \& Raftery(1995)}]{bayesfactor}
Kass, R.~E., \& Raftery, A.~E. 1995, Journal of the American Statistical
  Association, 90, 773

\bibitem[{{Lacki} {et~al.}(2014){Lacki}, {Horiuchi}, \&
  {Beacom}}]{2014ApJ...786...40L}
{Lacki}, B.~C., {Horiuchi}, S., \& {Beacom}, J.~F. 2014, \apj, 786, 40

\bibitem[{{Lisanti} {et~al.}(2016){Lisanti}, {Mishra-Sharma}, {Necib}, \&
  {Safdi}}]{2016ApJ...832..117L}
{Lisanti}, M., {Mishra-Sharma}, S., {Necib}, L., \& {Safdi}, B.~R. 2016, \apj,
  832, 117

\bibitem[{{Loredo} \& {Lamb}(1989)}]{1989NYASA.571..601L}
{Loredo}, T.~J., \& {Lamb}, D.~Q. 1989, Annals of the New York Academy of
  Sciences, 571, 601

\bibitem[{{Mattox} {et~al.}(1996){Mattox}, {Bertsch}, {Chiang}, {Dingus},
  {Digel}, {Esposito}, {Fierro}, {Hartman}, {Hunter}, {Kanbach}, {Kniffen},
  {Lin}, {Macomb}, {Mayer-Hasselwander}, {Michelson}, {von Montigny},
  {Mukherjee}, {Nolan}, {Ramanamurthy}, {Schneid}, {Sreekumar}, {Thompson}, \&
  {Willis}}]{1996ApJ...461..396M}
{Mattox}, J.~R., {Bertsch}, D.~L., {Chiang}, J., {et~al.} 1996, \apj, 461, 396

\bibitem[{{Narumoto} \& {Totani}(2006)}]{2006ApJ...643...81N}
{Narumoto}, T., \& {Totani}, T. 2006, \apj, 643, 81

\bibitem[{Neyman \& Pearson(1933)}]{neymanpeirson1933}
Neyman, J., \& Pearson, E.~S. 1933, Philosophical Transactions of the Royal
  Society of London. Series A, Containing Papers of a Mathematical or Physical
  Character, 231, 289

\bibitem[{{Nolan} {et~al.}(2012){Nolan}, {Abdo}, {Ackermann}, {Ajello},
  {Allafort}, {Antolini}, {Atwood}, {Axelsson}, {Baldini}, {Ballet},
  {Barbiellini}, {Bastieri}, {Bechtol}, {Belfiore}, {Bellazzini}, {Berenji},
  {Bignami}, {Blandford}, {Bloom}, {Bonamente}, {Bonnell}, {Borgland},
  {Bottacini}, {Bouvier}, {Brandt}, {Bregeon}, {Brigida}, {Bruel}, {Buehler},
  {Burnett}, {Buson}, {Caliandro}, {Cameron}, {Campana}, {Ca{\~n}adas},
  {Cannon}, {Caraveo}, {Casandjian}, {Cavazzuti}, {Ceccanti}, {Cecchi},
  {{\c{C}}elik}, {Charles}, {Chekhtman}, {Cheung}, {Chiang}, {Chipaux},
  {Ciprini}, {Claus}, {Cohen-Tanugi}, {Cominsky}, {Conrad}, {Corbet}, {Cutini},
  {D'Ammando}, {Davis}, {de Angelis}, {DeCesar}, {DeKlotz}, {De Luca}, {den
  Hartog}, {de Palma}, {Dermer}, {Digel}, {Silva}, {Drell}, {Drlica-Wagner},
  {Dubois}, {Dumora}, {Enoto}, {Escande}, {Fabiani}, {Falletti}, {Favuzzi},
  {Fegan}, {Ferrara}, {Focke}, {Fortin}, {Frailis}, {Fukazawa}, {Funk},
  {Fusco}, {Gargano}, {Gasparrini}, {Gehrels}, {Germani}, {Giebels},
  {Giglietto}, {Giommi}, {Giordano}, {Giroletti}, {Glanzman}, {Godfrey},
  {Grenier}, {Grondin}, {Grove}, {Guillemot}, {Guiriec}, {Gustafsson},
  {Hadasch}, {Hanabata}, {Harding}, {Hayashida}, {Hays}, {Hill}, {Horan},
  {Hou}, {Hughes}, {Iafrate}, {Itoh}, {J{\'o}hannesson}, {Johnson}, {Johnson},
  {Johnson}, {Johnson}, {Kamae}, {Katagiri}, {Kataoka}, {Katsuta}, {Kawai},
  {Kerr}, {Kn{\"o}dlseder}, {Kocevski}, {Kuss}, {Lande}, {Landriu},
  {Latronico}, {Lemoine-Goumard}, {Lionetto}, {Llena Garde}, {Longo},
  {Loparco}, {Lott}, {Lovellette}, {Lubrano}, {Madejski}, {Marelli}, {Massaro},
  {Mazziotta}, {McConville}, {McEnery}, {Mehault}, {Michelson}, {Minuti},
  {Mitthumsiri}, {Mizuno}, {Moiseev}, {Mongelli}, {Monte}, {Monzani},
  {Morselli}, {Moskalenko}, {Murgia}, {Nakamori}, {Naumann-Godo}, {Norris},
  {Nuss}, {Nymark}, {Ohno}, {Ohsugi}, {Okumura}, {Omodei}, {Orlando}, {Ormes},
  {Ozaki}, {Paneque}, {Panetta}, {Parent}, {Perkins}, {Pesce-Rollins},
  {Pierbattista}, {Pinchera}, {Piron}, {Pivato}, {Porter}, {Racusin},
  {Rain{\`o}}, {Rando}, {Razzano}, {Razzaque}, {Reimer}, {Reimer}, {Reposeur},
  {Ritz}, {Rochester}, {Romani}, {Roth}, {Rousseau}, {Ryde}, {Sadrozinski},
  {Salvetti}, {Sanchez}, {Saz Parkinson}, {Sbarra}, {Scargle}, {Schalk},
  {Sgr{\`o}}, {Shaw}, {Shrader}, {Siskind}, {Smith}, {Spandre}, {Spinelli},
  {Stephens}, {Strickman}, {Suson}, {Tajima}, {Takahashi}, {Takahashi},
  {Tanaka}, {Thayer}, {Thayer}, {Thompson}, {Tibaldo}, {Tibolla}, {Tinebra},
  {Tinivella}, {Torres}, {Tosti}, {Troja}, {Uchiyama}, {Vandenbroucke}, {Van
  Etten}, {Van Klaveren}, {Vasileiou}, {Vianello}, {Vitale}, {Waite},
  {Wallace}, {Wang}, {Werner}, {Winer}, {Wood}, {Wood}, {Wood}, {Yang}, \&
  {Zimmer}}]{2012ApJS..199...31N}
{Nolan}, P.~L., {Abdo}, A.~A., {Ackermann}, M., {et~al.} 2012, The
  Astrophysical Journal Supplement Series, 199, 31

\bibitem[{{Paliya} {et~al.}(2018){Paliya}, {Ajello}, {Rakshit}, {Mandal},
  {Stalin}, {Kaur}, \& {Hartmann}}]{2018ApJ...853L...2P}
{Paliya}, V.~S., {Ajello}, M., {Rakshit}, S., {et~al.} 2018, \apjl, 853, L2

\bibitem[{{Singal} {et~al.}(2012){Singal}, {Petrosian}, \&
  {Ajello}}]{2012ApJ...753...45S}
{Singal}, J., {Petrosian}, V., \& {Ajello}, M. 2012, \apj, 753, 45

\bibitem[{Wilks(1938)}]{wilks1938}
Wilks, S.~S. 1938, Ann. Math. Statist., 9, 60

\bibitem[{{Wood} {et~al.}(2017){Wood}, {Caputo}, {Charles}, {Di Mauro},
  {Magill}, {Perkins}, \& {Fermi-LAT Collaboration}}]{2017ICRC...35..824W}
{Wood}, M., {Caputo}, R., {Charles}, E., {et~al.} 2017, in International Cosmic
  Ray Conference, Vol. 301, 35th International Cosmic Ray Conference
  (ICRC2017), 824

\bibitem[{{Zechlin} {et~al.}(2016){Zechlin}, {Cuoco}, {Donato}, {Fornengo}, \&
  {Vittino}}]{2016ApJS..225...18Z}
{Zechlin}, H.-S., {Cuoco}, A., {Donato}, F., {Fornengo}, N., \& {Vittino}, A.
  2016, \apjs, 225, 18

\end{thebibliography}

\appendix
\section{Efficiency and logN-logS data}
Tables~\ref{tab:ph}-\ref{tab:ene} provide the efficiency ($\omega$) and {\it logN-logS} derived from our analysis and used in Figure~\ref{fig:eff}-\ref{fig:lognlogs}.  
\begin{table}[h!]
\centering
\scriptsize
\caption{Photon flux efficiency and {\it logN-logS}\label{tab:ph} associated with the survey at $|b|>20\degree$. The corresponding solid angle ($\Omega$) is $27143.61~\rm deg^2$.}
\begin{tabular}{ c c c }
 \hline
 Flux (\phflux) & $\omega$ & $dN/dS^{\rm +err_{stat}+err_{syst}}_{\rm -err_{stat}-err_{syst}} (\rm ph^{-1}cm^2s)$ \\
 \hline
 $(9.05\pm1.77)\times10^{-11}$ & $2.07^{+1.60}_{-0.91}\times10^{-4}$ & $1.36^{+3.34+4.75}_{-1.30-1.36}\times10^{14}$  \\ 
 $(1.34\pm0.26)\times10^{-10}$ & $5.03^{+2.30}_{-1.64}\times10^{-4}$ & $1.13^{+1.22+3.57}_{-0.71-0.27}\times10^{14}$\\
 $(2.00\pm0.39)\times10^{-10}$ & $1.38^{+0.29}_{-0.28}\times10^{-3}$ & $6.43^{+3.73+8.98}_{-2.71-1.27}\times10^{13}$ \\
 $(2.97\pm5.83)\times10^{-10}$ & $4.55^{+0.55}_{-0.54}\times10^{-3}$ & $4.14^{+1.03+6.10}_{-1.01-2.16}\times10^{13}$ \\
 $(4.43\pm0.86)\times10^{-10}$ & $7.85\pm{+0.74}\times10^{-3}$       &$2.57^{+0.50+3.19}_{-0.49-1.64}\times10^{13}$ \\ 
 $(6.58\pm1.29)\times10^{-10}$ & $2.33\pm{+0.13}\times10^{-2}$       & $1.29^{+0.16+1.58}_{-0.16-1.02}\times10^{13}$ \\
 $(9.80\pm1.92)\times10^{-10}$ & $6.01\pm{+0.25}\times10^{-2}$       & $5.79^{+0.56+7.38}_{-0.56-4.73}\times10^{12}$ \\
 $(1.45\pm0.28)\times10^{-9}$  & $1.56\pm{+0.05}\times10^{-1}$       & $2.30^{+0.18+3.28}_{-0.18-1.91}\times10^{12}$\\
 $(2.16\pm0.42)\times10^{-9}$  & $3.93\pm{+0.12}\times10^{-1}$       & $8.11^{+0.55+1.22}_{-0.55-7.09}\times10^{11}$\\ 
 $(3.22\pm0.63)\times10^{-9}$  & $7.51\pm{+0.23}\times10^{-1}$       & $3.40^{+0.22+5.40}_{-0.22-3.09}\times10^{11}$\\ 
 $(4.79\pm0.93)\times10^{-9}$  & $1.05\pm{+0.03}$                    & $1.77^{+0.11+2.24}_{-0.11-1.73}\times10^{11}$\\ 
 $(7.13\pm1.39)\times10^{-9}$  & $1.22\pm{+0.04}$                    & $9.33^{+0.06+1.16}_{-0.06-9.85}\times10^{11}$\\
 $(1.06\pm0.20)\times10^{-8}$  & $1.38\pm{+0.06}$                    & $5.23^{+0.39+6.59}_{-0.39-4.91}\times10^{10}$ \\
 $(1.57\pm0.30)\times10^{-8}$  & $1.38\pm{+0.07}$                    & $2.15^{+0.20+2.51}_{-0.20-1.99}\times10^{10}$\\ 
 $(2.34\pm0.45)\times10^{-8}$  & $1.05\pm{+0.06}$                    &  $1.21^{+0.14+1.76}_{-0.14-1.19}\times10^{10}$ \\ 
 $(3.49\pm0.68)\times10^{-8}$  & $1.03\pm{+0.08}$                    & $6.63^{+0.88+8.07}_{-0.88-6.08}\times10^{9}$ \\  
 $(5.19\pm1.01)\times10^{-8}$  & $1.00\pm{+0.09}$                    & $2.94^{+0.49+3.76}_{-0.48-2.70}\times10^{9}$  \\ 
 $(7.72\pm1.51)\times10^{-8}$  & $1.01\pm{+0.11}$                    &  $9.13^{+2.12+1.51}_{-2.09-7.89}\times10^{8}$ \\
 $(1.14\pm0.22)\times10^{-7}$  & $1.07\pm{+0.17}$                    & $3.70^{+1.07+4.37}_{-1.04-2.72}\times10^{8}$\\   
 $(1.70\pm0.33)\times10^{-7}$  & $0.92\pm{+0.18}$                    & $1.29^{+0.69+1.79}_{-0.51-9.45}\times10^{8}$  \\ 
 $(2.54\pm0.49)\times10^{-7}$  & $1.00\pm{0.22}$                     & $8.03^{+4.38+1.37}_{-3.31-4.71}\times10^{7}$  \\
 $(3.77\pm0.74)\times10^{-7}$  & $0.96\pm{0.26}$                     & $3.51^{+2.57+4.53}_{-1.79-2.69}\times10^{7}$  \\
 $(5.62\pm1.10)\times10^{-7}$  & $1.00\pm{0.29}$                     & $4.53^{+0.12+2.67}_{-6.45-8.28}\times10^{6}$ \\
 $(8.36\pm1.63)\times10^{-7}$  & $0.78\pm{0.41}$                     & $3.88^{+9.26+1.03}_{-3.57-3.88}\times10^{6}$ \\  
\hline  
\end{tabular}
\end{table}

\begin{table}[h!]
\centering
\scriptsize
\caption{Energy flux efficiency and {\it logN-logS}\label{tab:ene} associated with the survey at $|b|>20\degree$. The corresponding solid angle ($\Omega$) is $27143.61~\rm deg^2$.}
\begin{tabular}{ c c c }
 \hline
 Flux (\ergflux) & $\omega$ & $dN/dS^{\rm +err_{stat}+err_{syst}}_{\rm -err_{stat}-err_{syst}}(\rm erg^{-1}cm^2s)$ \\
 \hline
$(1.38\pm0.19)\times10^{-12}$ & $(4.68\pm0.36)\times10^{-2}$ & $1.62^{+0.33+2.21}_{-0.32-1.23}\times10^{15}$\\ 
$(1.86\pm0.25)\times10^{-12}$ & $(5.15\pm0.17)\times10^{-1}$ & $7.93^{+0.61+11.4}_{-0.61-6.59}\times10^{14}$\\
$(2.51\pm0.34)\times10^{-12}$ & $1.17\pm0.0)$ & $5.55^{+0.31+8.24}_{-0.31-4.79}\times10^{14}$ \\
$(3.39\pm0.47)\times10^{-12}$ & $1.37\pm0.04$ & $3.34^{+0.19+4.33}_{-0.19-3.01}\times10^{14}$ \\
$(4.57\pm0.63)\times10^{-12}$ & $1.41\pm0.05$ & $2.20^{+0.14+2.50}_{-0.14-2.00}\times10^{14}$ \\
$(6.17\pm0.85)\times10^{-12}$ & $1.30\pm0.05$ & $1.39^{+0.10+1.58}_{-0.10-1.28}\times10^{14}$ \\
$(8.32\pm1.15)\times10^{-12}$ & $1.29\pm0.06$ & $6.78^{+0.61+8.77}_{-0.61-5.42}\times10^{13}$ \\
$(1.12\pm0.15)\times10^{-11}$ & $1.22\pm0.07$ & $3.93^{+0.41+4.35}_{-0.41-3.67}\times10^{13}$ \\
$(1.51\pm0.21)\times10^{-11}$ & $1.17\pm0.08$ & $2.36^{+0.28+3.09}_{-0.28-1.89}\times10^{13}$ \\
$(2.04\pm0.28)\times10^{-11}$ & $1.16\pm0.08$ & $1.43^{+0.18+1.62}_{-0.18-1.17}\times10^{13}$ \\
$(2.75\pm0.38)\times10^{-11}$ & $1.12\pm0.09$ & $7.82^{+1.20+11.4}_{-1.19-5.60}\times10^{12}$ \\
$(3.72\pm0.51)\times10^{-11}$ & $1.07\pm0.11$ & $3.63^{+0.70+4.75}_{-0.69-2.53}\times10^{12}$ \\
$(5.01\pm0.69)\times10^{-11}$ & $1.05\pm0.13$ & $1.98^{+0.45+2.38}_{-0.44-1.38}\times10^{12}$ \\
$(6.76\pm0.93)\times10^{-11}$ & $(9.01\pm1.25)\times10^{-1}$ & $1.18^{+0.31+1.30}_{-0.30-0.98}\times10^{12}$ \\              
$(9.12\pm1.26)\times10^{-11}$ & $(9.40\pm1.66)\times10^{-1}$ & $6.73^{+2.09+12.4}_{-2.04-5.66}\times10^{11}$ \\              
$(1.23\pm0.17)\times10^{-10}$ & $1.02\pm0.22$ & $1.43^{+1.02+1.96}_{-0.68-1.23}\times10^{11}$\\
$(1.66\pm0.23)\times10^{-10}$ & $(8.48\pm1.86)\times10^{-1}$ & $1.54^{+0.98+1.92}_{-0.69-1.07}\times10^{11}$\\              
$(2.24\pm0.31)\times10^{-10}$ & $1.09\pm0.31$ & $1.04^{+6.38+1.32}_{-0.48-0.89}\times10^{11}$ \\         
$(3.02\pm0.41)\times10^{-10}$ & $(8.82\pm3.20)\times10^{-1}$ & $5.41^{+4.73+6.50}_{-3.21-3.61}\times10^{10}$\\ 
$(4.07\pm0.56)\times10^{-10}$ & $1.06\pm0.37$ & $1.55\times10^{10}$~\tablenotemark{a}\\
$(5.50\pm0.76)\times10^{-10}$ & $(8.33\pm4.77)\times10^{-1}$ & $7.88^{+0.18+0.15}_{-7.32-5.25}\times10^{9}$\\      
$(7.41\pm1.03)\times10^{-10}$ & $(8.57\pm6.92)\times10^{-1}$ & $5.68^{+14.0+7.10}_{-5.62-2.84}\times10^{9}$\\
\hline  
\multicolumn{3}{l}{%
  \begin{minipage}{0.5\textwidth}
\tablenotetext{a}{This is an upper limit at $1\sigma$ confidence level.}
\end{minipage}}

\end{tabular}
\end{table}
\clearpage

\end{document}